\pdfoutput=1

\newcommand{\schedulerbase}{\gls{bfd}}
\newcommand{\schedulerfreq}{\gls{bcffs}}
\newcommand{\schedulermigr}{\gls{bcf}}

\newcommand{\ensavingsmax}{$14.57\%$}
\newcommand{\ensavingsfreq}{$14.13\%$}

\newcommand{\ensavingsmaxintel}{$10.06\%$}
\newcommand{\ensavingsfreqintel}{$9.86\%$}

\newcommand{\vmnumsimulation}{2k}
\newcommand{\pmnumsimulation}{2k}

\documentclass[10pt,journal,compsoc]{IEEEtran}

\usepackage[T1]{fontenc} 
\usepackage[utf8]{inputenc}
\usepackage{url}
\usepackage{graphicx}
\usepackage{adjustbox}
\DeclareGraphicsExtensions{.pdf,.png,.jpg}
\graphicspath{{./img/}{./img2/}}
\usepackage{algpseudocode}
\usepackage{flushend} 
\usepackage{amsmath}
\usepackage{float}
\usepackage{amssymb}
\usepackage{glossaries}
\usepackage[super]{nth}
\usepackage{mathptmx}
\usepackage{todonotes}
\newcommand{\coe}{$CO_{2}e$}

\newcommand{\eqtopmargin}{-0.1cm}

\newcommand{\figtopmargin}{-0cm}
\newcommand{\figbottommargin}{-0cm}
\newcommand{\figcaptionmargin}{-0cm}

\newacronym{coe}{\coe}{$CO_{2}$ equivalent}
\newacronym{ewma}{EWMA}{exponentially weighted moving average}
\newacronym{dvfs}{DVFS}{dynamic voltage \& frequency scaling}
\newacronym{vm}{VM}{virtual machine}
\newacronym{era}{ERA}{energy reduction assets}
\newacronym{api}{API}{application programming interface}
\newacronym{os}{OS}{operating system}
\newacronym{rtp}{RTP}{real-time pricing}
\newacronym{qos}{QoS}{quality of service}
\newacronym{sla}{SLA}{service level agreement}
\newacronym{rtep}{RTEP}{real-time electricity pricing}
\newacronym{iaas}{IaaS}{infrastructure as a service}
\newacronym{pm}{PM}{physical machine}
\newacronym{pue}{PUE}{power usage efficiency}
\newacronym{cue}{CUE}{carbon usage effectiveness}
\newacronym{cef}{CEF}{carbon emission factor}
\newacronym{hpc}{HPC}{high-performance computing}
\newacronym{db}{DB}{database}
\newacronym{dc}{DC}{data center}
\newacronym{oltp}{OLTP}{online transaction processing}
\newacronym{mse}{MSE}{mean squared error}
\newacronym{ga}{GA}{genetic algorithm}
\newacronym{arima}{ARIMA}{autoregressive integrated moving average}
\newacronym{ses}{SES}{simple exponential smoothing}
\newacronym{bcf}{BCF}{best cost fit}
\newacronym{bfd}{BFD}{best fit decreasing}
\newacronym{iot}{IoT}{Internet of Things}
\newacronym{bcffs}{BCFFS}{Best Cost Fit Frequency Scaling}
\newacronym{fps}{FPS}{frames per second}
\newacronym{hevc}{HEVC}{high efficiency video coding}

\usepackage{tikz}

\usetikzlibrary{fit,positioning}
\usetikzlibrary{arrows}
\usetikzlibrary{calc}

\usepackage{booktabs}
\usepackage{algpseudocode}
\usepackage{algorithm}
\usepackage{varwidth}
\usepackage{relsize}

\ifCLASSOPTIONcompsoc
  \usepackage[nocompress]{cite}
\else
  \usepackage{cite}
\fi

\begin{document}

\title{Performance-Based Pricing in Multi-Core Geo-Distributed Cloud Computing}
 	

\author{
	Dražen~Lučanin,
	Ilia~Pietri,
	Simon Holmbacka,
	Ivona~Brandic,
	Johan~Lilius
	and~Rizos~Sakellariou
	%
  	\thanks{Manuscript received January 31, 2016.}
  	\thanks{The first three authors contributed equally to this work.}
	\thanks{D. Lučanin and I. Brandic are with the Vienna University of Technology, Vienna, Austria. 
    E-mail: \{drazen.lucanin, ivona.brandic\}@tuwien.ac.at}
    \thanks{I. Pietri and R. Sakellariou are with the University of Manchester, Manchester, United Kingdom. 
    E-mail: \{pietrii, rizos\}@cs.man.ac.uk}
    \thanks{S. Holmbacka and J. Lilius are with the Åbo Akademi University, Turku, Finland. 
    E-mail: \{sholmbac,jolilius\}@abo.fi}
}

\markboth{IEEE Transactions on Cloud Computing}%
{Shell \MakeLowercase{\textit{et al.}}: Bare Demo of IEEEtran.cls for Computer Society Journals}

\IEEEtitleabstractindextext{%
\begin{abstract}
New pricing policies are emerging where cloud providers charge resource provisioning based on the allocated CPU frequencies.\
As a result, resources are offered to users as combinations of different performance levels and prices which can be configured at runtime.\
With such new pricing schemes\ 
and the increasing energy costs in data centres, balancing energy savings with performance and revenue losses is a challenging problem for cloud providers.\
CPU frequency scaling can be used to reduce power dissipation, but also impacts \gls{vm} performance and therefore revenue.\
In this paper, we firstly propose a non-linear power model that estimates power dissipation of a multi-core \gls{pm} and secondly a pricing model that adjusts the pricing based on the \gls{vm}'s CPU-boundedness characteristics.\
Finally, we present a cloud controller that uses these models to allocate \glspl{vm}\
and scale CPU frequencies of the \glspl{pm}\ 
to achieve energy cost savings that exceed service revenue losses. We evaluate the proposed approach using simulations with realistic \gls{vm} workloads, electricity price and temperature traces and estimate energy savings of up to \ensavingsmax{}.

\end{abstract}

}

\maketitle




\section{Introduction}

Infrastructure-as-a-service is the fastest growing segment of the cloud services market
according to a 2013 Gartner report \cite{_gartner_2013}
(with a 42.4\% annual growth rate) and new \textit{performance-based pricing} models are being introduced by cloud providers like ElasticHosts \cite{elastichosts} and CloudSigma \cite{cloudsigma} that radically change the cloud computing revenue models and 
require new cloud control approaches.\
In performance-based pricing, the cost of \gls{vm} provisioning is based on the selected CPU frequency along with the allocated amount of RAM and the use of other resources,
and the \gls{vm}'s performance can be modified by choosing from a range of CPU frequencies and matching prices -- even at runtime.\
This approach mainly targets users who potentially need a lower performance level and would prefer a proportionally scaled price.

On the other hand, cloud providers are interested in minimizing the energy costs required to operate data centres. Data centers correspond to over 2\% of the total electricity consumed in the US \cite{jonathan_koomey_growth_2011} and the ICT sector's global $CO_{2}$ emissions have surpassed those of aviation \cite{_gartner_2007}.
The problem of energy efficiency becomes even more challenging in geographically-distributed data centres where the energy costs are influenced by dynamic local factors, such as real-time electricity prices \cite{weron_modeling_2006} and temperature-dependent cooling \cite{xu_temperature_2013}. 
We refer to these factors as geotemporal inputs. 
A power management action commonly used to reduce energy costs is CPU frequency scaling \cite{miyoshi2002critical}.\
Potential energy savings can be estimated based on geotemporal inputs.\
CPU frequency scaling actions, however, also cause service revenue losses in performance-based pricing.
Hence, the goal of this paper is to
find cloud control approaches that balance energy savings and service revenue losses caused by CPU frequency scaling in performance-based pricing.

Cloud control solutions relying on CPU frequency scaling\
exist, e.g. \cite{von2009power,shi2011towards}, but only consider fixed \gls{vm} pricing where the trade-offs of energy savings and performance-based \gls{vm} pricing are not considered.\
The currently existing clock frequency governors available at the operating system level that adjust CPU clock frequency according to workload changes have proven to be inefficient in responding to the required \gls{vm} performance level \cite{holmbacka2015energy,holmbacka2014energy}.

Other open challenges are that modern physical machines have multiple CPU cores with complex utilisation-to-power-dissipation models. Additionally, besides the traditional CPU architecture like Intel's, smartphone-technology-based ARM CPU architectures are emerging in large scale cloud platforms \cite{rajovic2013supercomputing,francesquini2015benchmark} with significantly different power models. Finally,
the performance impact of the clock frequency may also vary between different workloads \cite{spiliopoulos2014power}.\
For example, CPU-bound workloads are more sensitive to the reduction in frequency, while the performance of I/O-bound workloads is less affected. The sensitivity of the workload to CPU frequency reduction is called the workload's CPU-boundedness $\beta$ following the approach in \cite{etinski2010optimizing}.\
A cloud control solution has to model and consider such environments to be of practical relevance.

In this paper, we introduce a compound cloud control model which considers all the mentioned factors representative of modern cloud systems. We combine:\
(1) Realistic power modelling accounting for multi-core, Intel and ARM architectures;\
(2) Energy cost calculation based on geotemporal inputs;\
(3) Performance-based \gls{vm} pricing;\
(4) Variable \gls{vm} CPU-boundedness that determines the performance impact of CPU frequency scaling.



To describe real-world power dissipation behaviour, we developed a non-linear power model based on real experiments performed on multi-core Intel and ARM CPU architectures representative of modern data center infrastructures \cite{rajovic2013supercomputing,francesquini2015benchmark}. As we show, on such power models, traditional race-to-idle approaches \cite{sasaki2013model,seeker2014energy} are no longer valid, which also motivates the cloud control method we introduce in this paper.

To tackle varying \gls{vm} workloads,\
we propose a novel \textit{perceived-performance pricing} scheme for determining the \gls{vm} price based on the application-level performance.\
This scheme allows energy-aware cloud control that treats \glspl{vm} differently based on the actual impact that CPU frequency scaling will have on their workload performance, considering their measured CPU-boundedness.

To address the data center energy consumption and performance-based \gls{vm} revenue trade-offs,\
we introduce the \schedulerfreq{} cloud controller.\
The controller was adapted for new power models and pricing data from our initial work in \cite{lucanin_cloud_2015}.\
The controller we propose uses our multi-core power model for ARM and Intel CPU architectures in an energy calculation method that factors in geotemporal inputs from multiple geo-distributed data centers.\
To account for performance-based \gls{vm} pricing, ElasticHosts \cite{elastichosts} and CloudSigma \cite{cloudsigma} price data was used to model their behaviour and precisely compute the effects of each CPU frequency level.\
The \schedulerfreq{} cloud controller then combines both models in a two-phase algorithm, where firstly \glspl{vm} are allocated between geo-distributed data centers and subsequently CPU frequencies are set for each \gls{pm} where energy savings exceed service revenue losses.\

The controller and the models were mapped onto the Philharmonic simulator \cite{lucanin2014energy}.
Simulations with a wide range of scenarios are used to estimate the energy savings and service revenue stemming from our cloud control approach.\ 
The results obtained by the \schedulerfreq{} cloud controller are compared and evaluated using two baseline controllers \cite{beloglazov_energy-aware_2012} and historical traces of real-time electricity prices~\cite{alfeld_toward_2012} and temperatures~\cite{lucanin2014energy}. The \gls{vm} CPU-boundedness values used in the simulation are distributed according to the PlanetLab~\cite{planetlab} dataset of \gls{vm} CPU usage. The results indicate that energy savings up to \ensavingsmax{} without significant service revenue reductions can be achieved using the \schedulerfreq{} cloud controller.

We structure the paper by first examining the related work in the following section. We then introduce in more details the challenges of frequency scaling in multi-core computers and the inefficiencies of existing control approaches in Section~\ref{sec:challenges}. This serves as a motivation for our detailed multi-core power model for Intel and ARM CPU architectures in Section~\ref{sec:power_model}. We then highlight the economical aspects of frequency scaling in cloud computing by explaining emerging \gls{vm} pricing schemes and propose our own perceived-performance pricing scheme in Section~\ref{sec:pricing_model}. We combine the power and pricing models to devise a cloud controller that geographically distributes \gls{vm}s and applies frequency scaling on \gls{pm}s in Section~\ref{sec:scheduler}. In Section~\ref{sec:evaluation}, we present the evaluation methodology and comment on the most significant obtained results. Finally, we conclude the paper in Section~\ref{sec:conclusion}.



\section{Related Work}

Analysing geotemporal inputs to optimise distributed systems was studied previously. Network routing in content delivery networks is adapted for \gls{rtep} in~\cite{qureshi_cutting_2009} with reported savings of up to 40\% of total electricity costs. Job placement based on geotemporal inputs for map-reduce jobs is researched in \cite{buchbinder_online_2011} and for computational grids based on both \gls{rtep} and cooling in \cite{guler_cutting_2013,liu_renewable_2012}. Geotemporal inputs as a basis for scaling CPU frequencies or as a counter-balance to performance-based pricing has not been researched prior to our work.

Frequency scaling is the focus in many studies with the aim to reduce power consumption by decreasing the CPU frequency \cite{etinski2010optimizing,von2009power,wu2014green}. The cloud scheduler in \cite{wu2014green} sorts and allocates the incoming jobs to \gls{vm}s based on the user \gls{sla}s. The minimum resource requirements are allocated to each \gls{vm} and the CPU frequencies of the \gls{pm}s with low load are reduced so that resource wastage is minimised without affecting the performance of the executing jobs. In \cite{von2009power}, the proposed scheduler allocates the queued \gls{vm}s to \gls{pm}s, while reducing the CPU frequencies at runtime so that \gls{vm} performance requirements can be met, preferring \gls{pm}s that operate at lower frequencies.
As opposed to related work, our proposed controller scales the CPU frequencies taking into account the workload CPU-boundedness while controlling the impact of frequency reduction on the provider's profit.
The impact of frequency scaling on workload performance has been investigated in many studies \cite{etinski2010optimizing,miyoshi2002critical,freeh2007analyzing}. 
In \cite{miyoshi2002critical}, the authors investigate the power-performance characteristics of systems with frequency scaling capabilities and introduce a metric to determine energy-efficient performance points to operate the system. This is also the focus in \cite{freeh2007analyzing}, investigating the impact of frequency scaling on workload performance for different HPC workloads in order to achieve energy-performance trade-offs.

Although \gls{dvfs} cloud controllers have been proposed before \cite{hsuan2013cloud,zhuo2014cloud,ioannou2011cloud,alnowiser2014cloud},\
our adaptive approach scales the operating frequencies based on the \gls{vm} CPU-boundedness and the impact frequency reduction has on the provider's gross profit under performance-based pricing.
Also, in most papers multi-core modelling is not considered or is simplified as a linear combination of the number of cores used.
The clock frequency of the systems used in cloud platforms are usually modelled according to its dynamic power as a product of the clock frequency and the core voltage. In contrast to such systems, we focus on adopting a more accurate multi-core power model; still simple enough to be integrated in real systems. 
The model accounts for real-world influences more accurately such as the heat dissipation influencing the static power significantly \cite{hallis2013power}.


The literature has shown many power models and approaches to model the power dissipation in computer systems \cite{martinez2010model,rauber2012model,tudor2012model,sasaki2013model,cupertino2014model,shao2013model}.
Most models are constructed bottom-up from physical characteristics on top of which practical aspects such as frequency scaling is applied.
The dynamic power dissipation is expressed in many examples \cite{cho2010model,shen2012model,bharathwaj2013model} as the relation $f \cdot v^\alpha$ where $f$ is the clock frequency, $v$ is the core voltage and $\alpha$ is a constant used to comply with the real platform as close as possible.

As the dynamic power dissipation can be expressed accurately with this simple bottom-up formula,
the ever growing static power proves more difficult.
The leakage current causing the static power is expressed in \cite{kim2003power} as a relationship between transistor gate width, thermal voltage and architectural parameters such as the insulation material.
Moreover, leakage is also caused by electron tunnelling through the insulator.
This means that a bottom-up modelling of static power is significantly more difficult.
We instead used a top-down view of the power model, purely based on real-world experiments,
which provides a more realistic view of the complete system including cores, buses, memories, temperature, operating system influence and other software.




\section{Challenges of Multi-Core\\ Frequency Scaling}
\label{sec:challenges}

In this section we introduce the main challenges inherent to multi-core frequency scaling of \gls{pm}s with multiple CPU cores and motivate our power model and subsequently frequency control approach. We show that neither operating system CPU governors, nor traditional race-to-idle approaches\
provide optimal energy efficiency because of inaccurate decision making.



\subsection{Limitations of Current Frequency Governors}
The currently used power management system in Linux operating systems is handled by the \textit{frequency governors},
which alter the CPU clock frequency based on a predefined policy.
Even though the intention is to reduce the clock frequency when performance is not needed,
the approach suffers from limitations.

\begin{figure}[!t]
\centering
\includegraphics[width=7cm]{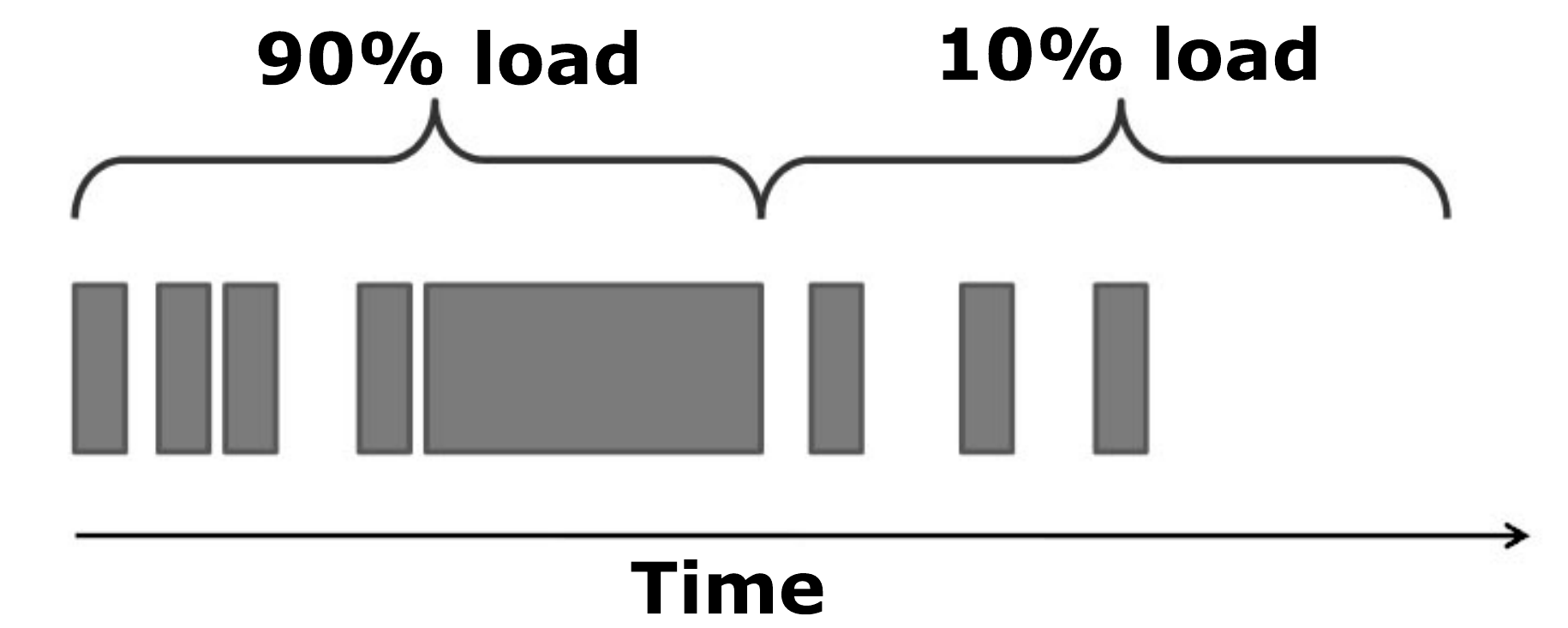}
\caption{Load calculated as a ratio between active and idle CPU for a defined window -- dark squares indicate an active CPU.}
\label{fig:loadwindow}
\end{figure}

The metric used to determine the clock frequency in the governors is the system \textit{workload}.
The workload is expressed as a ratio between an active CPU and an idle CPU over a given time window,
which is illustrated in Fig.~\ref{fig:loadwindow} as two time windows: one with 90\% load and one with 10\% load.

Workload, however, does not represent the performance, or "real work done" of an application, but mainly the activity level of the CPU.
This means that as long as the CPU is loaded, the performance requirement is recognised as insufficient and the clock frequency is increased to the maximum even though the \textit{actual} performance is sufficient on a moderate clock frequency.

\subsection{Energy Inefficient Execution}
Using workload as the metric for power management decisions often results in race-to-idle scenarios \cite{sasaki2013model,seeker2014energy}, in which the workload is executed as fast as possible in order to obtain an idle system.
This execution principle was considered an energy efficient method of executing workload in previous generation single-core microprocessors, because the minimisation of the execution time caused minimal energy consumption.

\begin{figure}[t]
\centering
\includegraphics[width=8cm]{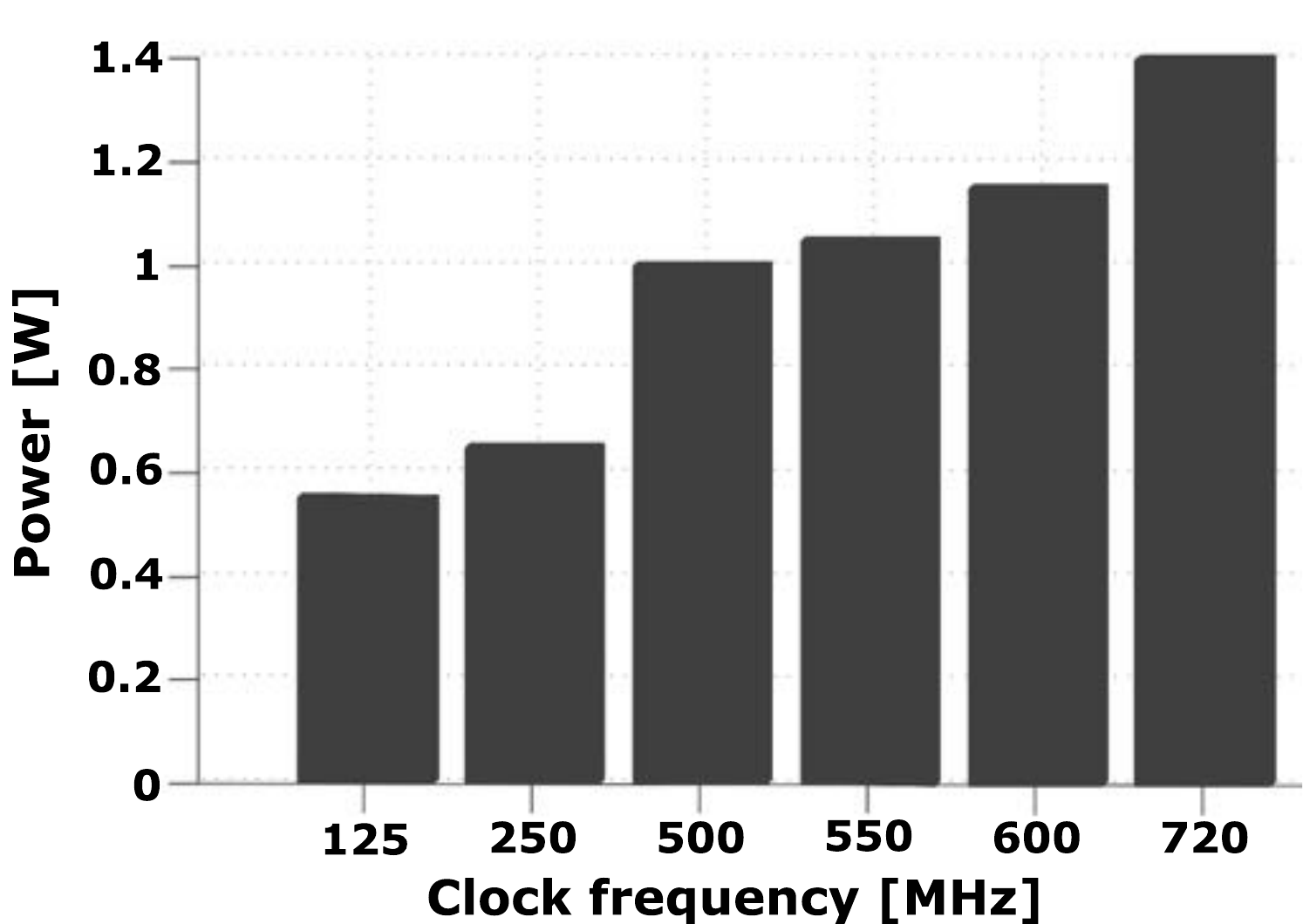}
\caption{Power dissipation for a Cortex-A8 CPU using different clock frequencies.}
\label{fig:cortexa8}
\end{figure}

This is demonstrated in Fig.~\ref{fig:cortexa8}, which shows the total power dissipation for a single-core ARM Cortex-A8 processor using different clock frequencies.
As seen in the figure, the highest clock frequency (720 MHz) results in roughly 1.4W of power dissipation.
When scaling down the frequency roughly 3x (250 MHz), the power dissipation is only reduced by 2x (0.7W), which means that the total energy consumption may be lower when executing at a higher clock frequency.

\begin{figure}[!b]
\centering
\includegraphics[width=1.0\columnwidth]{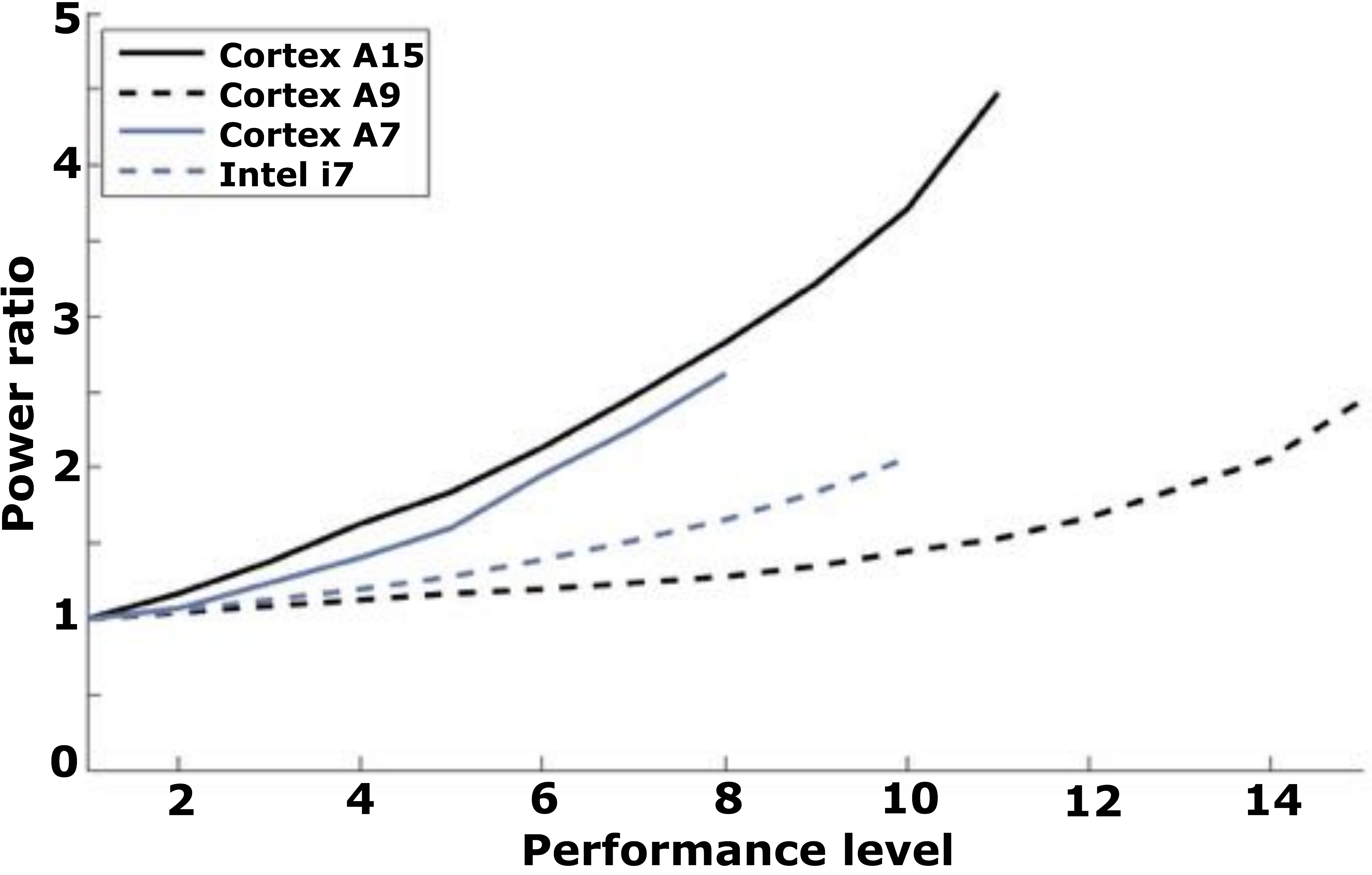}
\caption{Performance to power ratio for different platforms.}
\label{fig:profile}
\end{figure}

However, using more recent microprocessors with higher clock frequencies and multiple cores, the power dissipation has increased exponentially -- this has reduced the energy efficiency of the race-to-idle principle because the cost in power is greater than the savings in execution time \cite{holmbacka2014energy,holmbacka2015energy,seeker2014energy}.
Fig.~\ref{fig:profile} shows the relative performance-to-power ratio of four different modern platforms.
All of the four platforms show an exponential profile, which means that the power dissipation required to operate on the highest clock frequencies is higher than the relative performance gain of the platform.
The race-to-idle principle should therefore not be used for energy efficient execution.

\subsection{Energy Efficient Execution}
In order to not race-to-idle, a \textit{performance driven execution} should be used instead of a workload driven execution \cite{holmbacka2014energy,holmbacka2015energy,seeker2014energy}.
By monitoring the actual performance of an application and adjusting the clock frequency accordingly,
the energy efficiency can be improved. 

Benchmarks were executed on a quad-core ARM platform with an Exynos 5410 SoC using the default \textit{ondemand} governor and the modified performance driven power manager.
By using the ondemand governor, the decoder decodes the frames as quickly as possible
since the decoding task increases the workload, and the clock frequency is consequently increased.
As the frame buffer is filled, the decoder is idle until the frame buffer is emptied by the video display.
By instead decoding at the same frame rate as the video display is using (25 FPS), the clock frequency can be reduced to an intermediate clock frequency for the whole execution while still providing the required video quality.

\begin{figure}[t]
\centering
\includegraphics[width=1.0\columnwidth]{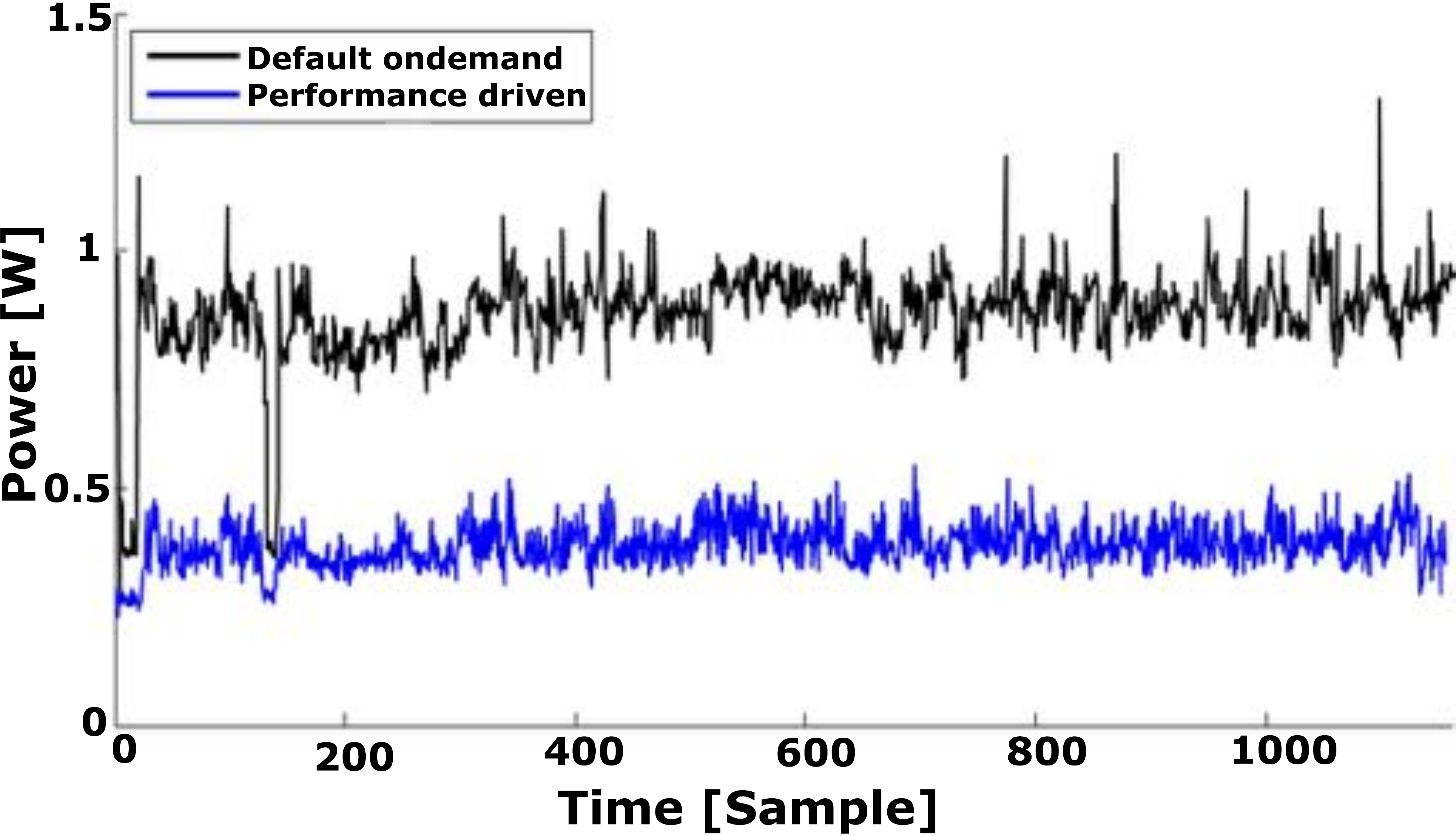}
\caption{Experiments with video decoding using the ondemand governor and performance driven clock frequency scaling on an Exynos 5410 platform \cite{holmbacka2015energy}.}
\label{fig:mplayer}
\end{figure}

The power dissipation was measured by internal power sensors for both power managers and the result is shown in Fig.~\ref{fig:mplayer} (more details can be found in \cite{holmbacka2015energy}).
The power dissipation of executing the video decoder using the ondemand governor is shown as the upper, black line and the performance driven power manager is shown as the lower, blue line.
Since the standard ondemand governor increases the clock frequency while workload is present, most of the execution demands the highest clock frequencies, which causes excess power dissipation as seen in Fig.~\ref{fig:mplayer} (upper, black line).
By matching the decoding framerate to the output framerate (of 25 \gls{fps}),
lower clock frequencies are providing enough performance to decode the frames at the intended phase of 25 \gls{fps},
and the power is significantly reduced seen in Fig.~\ref{fig:mplayer} (lower, blue line).

We intend to bring the performance driven clock frequency scaling using multi-core hardware from CPU level to the cloud level consisting of many parallel machines.
The difference is a much more diverse execution platform with additional parameters such as VM migration, network I/O and variable electrical cost models.
A power model capable of reflecting such details is needed to get an accurate cost model of the cloud system.


\section{Multi-Core Power Model}
\label{sec:power_model}


In this section we show how we modelled the behaviour of multi-core CPU power dissipation, accounting for both the CPU frequency and the number of active cores for generic multi-core systems. 
Such a power model allows us to determine what performance level to execute at, depending on the performance requirements, which is one of the key parts of our cloud controller.

As the power characteristics of modern multi-core CPUs are highly non-linear \cite{holmbacka2014energy}, a non-linear model should be created to accommodate as accurately as possible to the real-world power dissipation.
The model should also not be computationally heavy to introduce unnecessary overhead.

\subsection{ARM and Intel Architectures in the Cloud}
\label{sec:mont_blanc}

Aside from the popular Intel architecture used as a typical server platform,
the architecture based on ARM processors made popular through wide usage in smartphones is currently also being investigated for use in servers. ARM processors are much more energy efficient than Intel processors, though their maximum CPU frequency capacity is lower, potentially increasing the necessary number of servers and therefore the communication overhead. The Mont Blanc EU project \cite{rajovic2013supercomputing,francesquini2015benchmark} was devoted to determine whether this approach is valid for large scale cloud platforms. Companies like Calxeda already ship ARM based server machines and Lenovo is pushing its NextScale \cite{shah2015platform} platform with the motivation to increase the performance-per-watt ratio by focusing on possibly more energy efficient architectures.

We therefore included the ARM architecture in our evaluation as a viable candidate for investigating the effects of performance-based frequency scaling in order to provide a comparison to the Intel architecture.

\subsection{Power and Energy Consumption Model}
\label{sec:power_model_arm}

\begin{figure}[t]
\centering
\includegraphics[width=1.0\columnwidth]{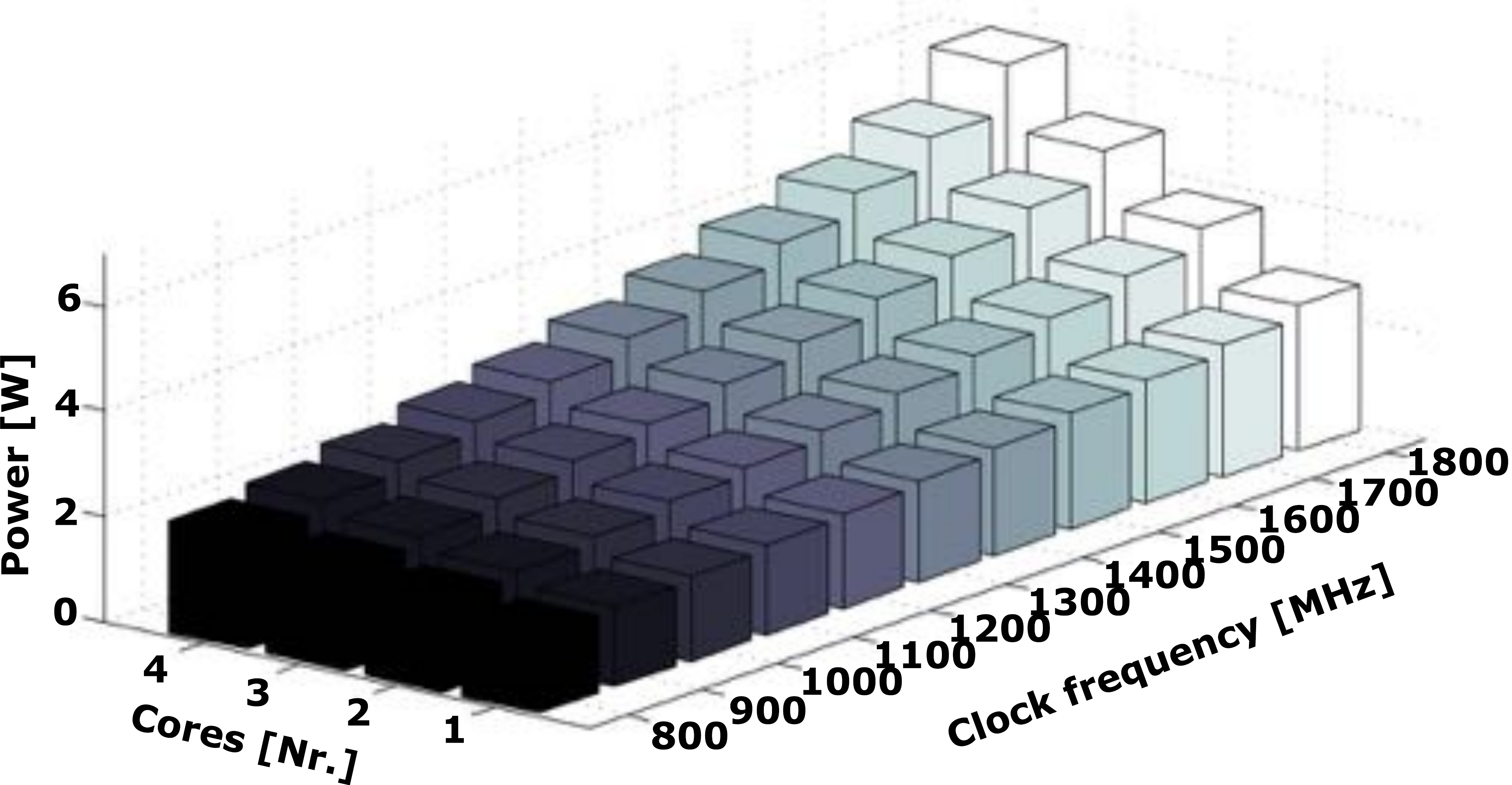}
\caption{A top-down model of a quad-core ARM Cortex-A15 CPU. The figure shows the power dissipation of the CPU during full load using different configurations of clock frequency and number of cores.}
\label{fig:odroidpower}
\end{figure}

For our cloud controller, we created an ARM and an Intel power consumption model.\ 
The ARM model was created by reading internal power sensors on an Exynos 5410 board,
and the Intel model by using an external measurement device connected directly to the ATX socket on the motherboard.
Both models were used in the evaluation, but we describe our modelling procedure with a higher focus on the ARM model for brevity.
The same procedure we describe in this section is also applicable to the Intel model (or any other derived model).

We used a similar methodology as the work in \cite{holmbacka2014energy} to derive the power model, where the model was created by stressing the system to full capacity using all combinations of the clock frequencies and all number of cores.
Our power measurements for stressing the ARM board is shown in Fig.~\ref{fig:odroidpower}.
Naturally, more cores and a higher clock frequency cause a higher power dissipation.

\begin{figure}[!t]
\centering
\includegraphics[width=1.0\columnwidth]{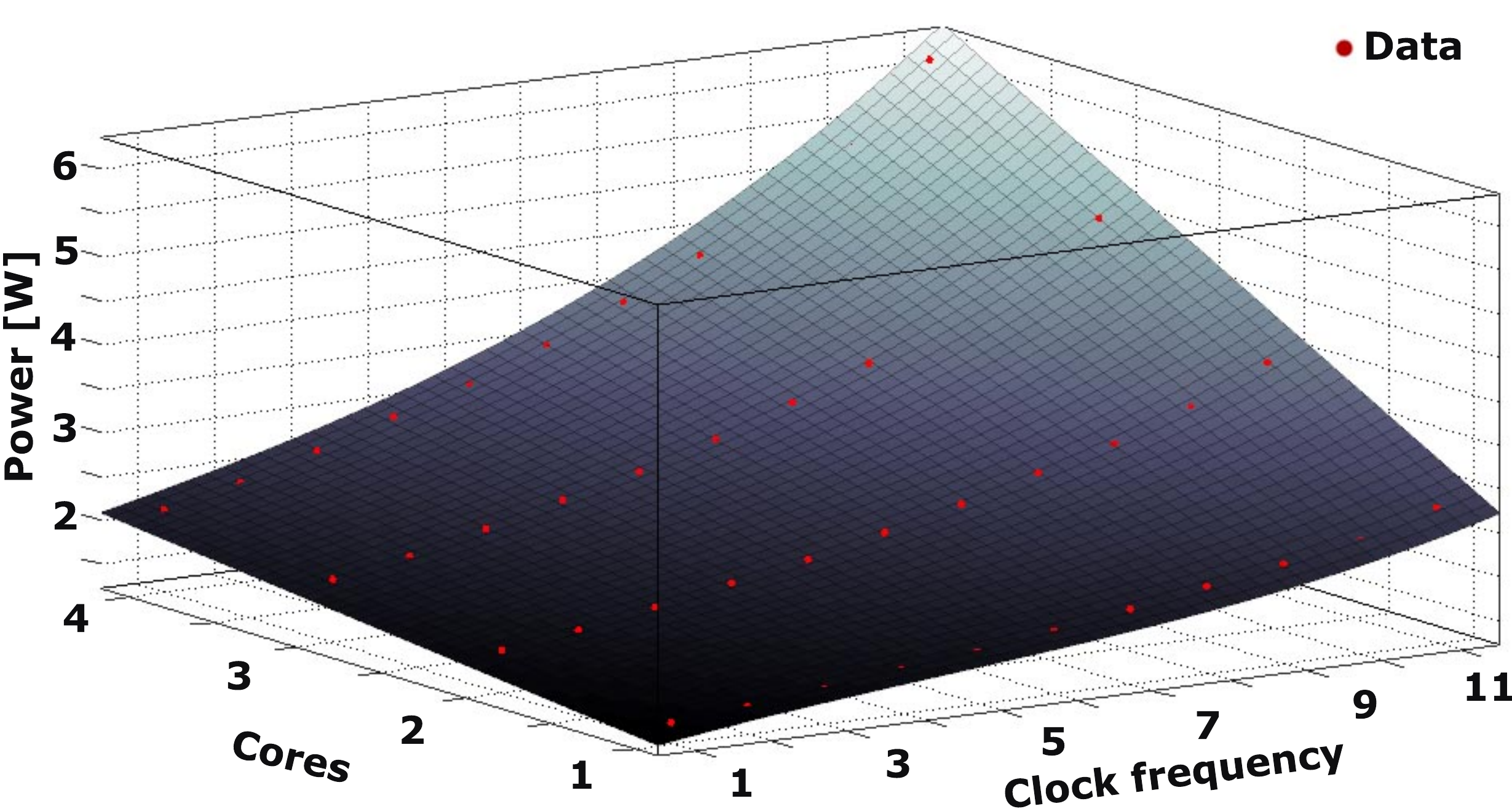}
\caption{A mathematical representation of the power values in Fig.~\ref{fig:odroidpower}
using surface fitting methods.}
\label{fig:planefitting}
\end{figure}

The power measurements were then used as basis for a two dimensional plane fitting algorithm, \
in order to build a mathematical expression of the multi-core system and its power dissipation.
We used a least-squares algorithm \cite{lawson1987optimization} provided in Matlab to obtain the polynomial of the form:
\begin{align}\label{eq:ppeak}
P_f(q,c)=p_{00}+p_{10}q+p_{01}c+p_{20}q^2+p_{11}qc+p_{30}q^3+p_{21}q^2c
\end{align}
which is a function of the clock frequency ($q$) and number of cores ($c$) used.
Fig.~\ref{fig:planefitting} shows the analytical representation of the power dissipation and the data points obtained from Fig.~\ref{fig:odroidpower} for the ARM platform.
The clock frequency and the number of cores used are represented as discrete steps from 1 to 11 and from 1 to 4 respectively.
The plane shown in Fig.~\ref{fig:planefitting} was fitted to the data values using the obtained parameters shown in Table~\ref{tab:parameters}.
The same method was used for the Intel platform, and other parameters were then obtained.


\begin{table}[!b]
 \caption{\small Power model coefficients.}
  \begin{center}
  \begin{tabular}{ | c | c | c | c | c | c | c|}
  \hline
  $p_{00}$  & $p_{01}$ & $p_{10}$ & $p_{11}$ & $p_{20}$ & $p_{21}$ & $p_{30}$   \\ \hline
  \footnotesize 1.318  &\footnotesize 0.03559 &\footnotesize 0.2243 &\footnotesize -0.00318 &\footnotesize 0.03137 &\footnotesize 0.000438 &\footnotesize 0.00711  \\ \hline
  \end{tabular}
\label{tab:parameters}
\end{center}
\end{table}

By comparing the model to the measurements, a maximum deviation of 18.7\% was obtained and an average deviation of 6.4\% compared to the experimental data, which we considered feasible for our cloud controller evaluation.

The idle power was modelled similarly to Eq.~\ref{eq:ppeak}, but without the core component $c$ as:
\begin{align}\label{eq:pfixed}
P_{idle}=p_{00}+p_{10}q+p_{20}q^2+p_{30}q^3
\end{align}
where $q$ is the clock frequency step and the p-parameters are identical to the fitted parameters in Table~\ref{tab:parameters}.

\subsection{I/O based power dissipation} \label{section:berserk}






The power dissipation of a \gls{pm} varies depending on the CPU utilisation of the machine,
which is dependent of the I/O usage of the workload. 
A \emph{CPU-boundedness parameter ($\beta$)} is therefore used to model the portion of the execution which consists of low intensity I/O operations.
The parameter may range between 0 and 1 to represent workloads of different CPU-boundedness properties with values close to 0 corresponding to I/O intensive workloads and values close to 1 corresponding to CPU-bound workloads \cite{etinski2010optimizing}. 
The value of $\beta$ normalises the ratio between low intensity I/O bound and high intensity CPU bound instructions in the workload.

The power model was therefore extended to account for the CPU utilisation based on the \gls{vm} CPU-boundedness of the executing workload. To do so, the CPU utilisation $u$ is expressed as:
 \begin{align}\label{eq:utilisation}
 u=\sum\limits_{core}\frac{\gamma_{core}}{cores_{active}}
 \end{align}
where $\gamma_{core}$ is a power ratio depending on the \gls{vm} CPU-boundedness $\beta$ and $cores_{active}$ represents the number of the currently used cores of the \gls{pm}.


Similarly to the basic power model (Eq.~\ref{eq:ppeak}), the power dissipation was evaluated on the same platform with different $\beta$ in order to train the model.
The experiments were run using the \textit{Berserk} benchmark on a single active core (to avoid any lack of scalability from using multiple cores and get the pure effect of only the I/O).
The Berserk benchmark is an open source application\footnote{https://github.com/philharmonic/berserk} that we developed for stressing the CPU cores at various CPU-boundedness ratios $\beta$. The workload itself is a CPU-intensive task -- repeated recursive calculation of Fibonacci numbers executed on all available CPU cores. By passing different $\beta$ parameters to the benchmark, proportional ratios of the workload are deferred to a remote server, making the work more or less CPU-bound for monitoring purposes. For example for a value of $\beta$ equal to 0, all the work is sent to and received from a remote server via the network, making the task fully I/O-bound. For a value of $\beta$ equal to 1, all the work is executed locally, resulting in a CPU-bound task.

\begin{figure}[t]
\centering
\includegraphics[width=1.0\columnwidth]{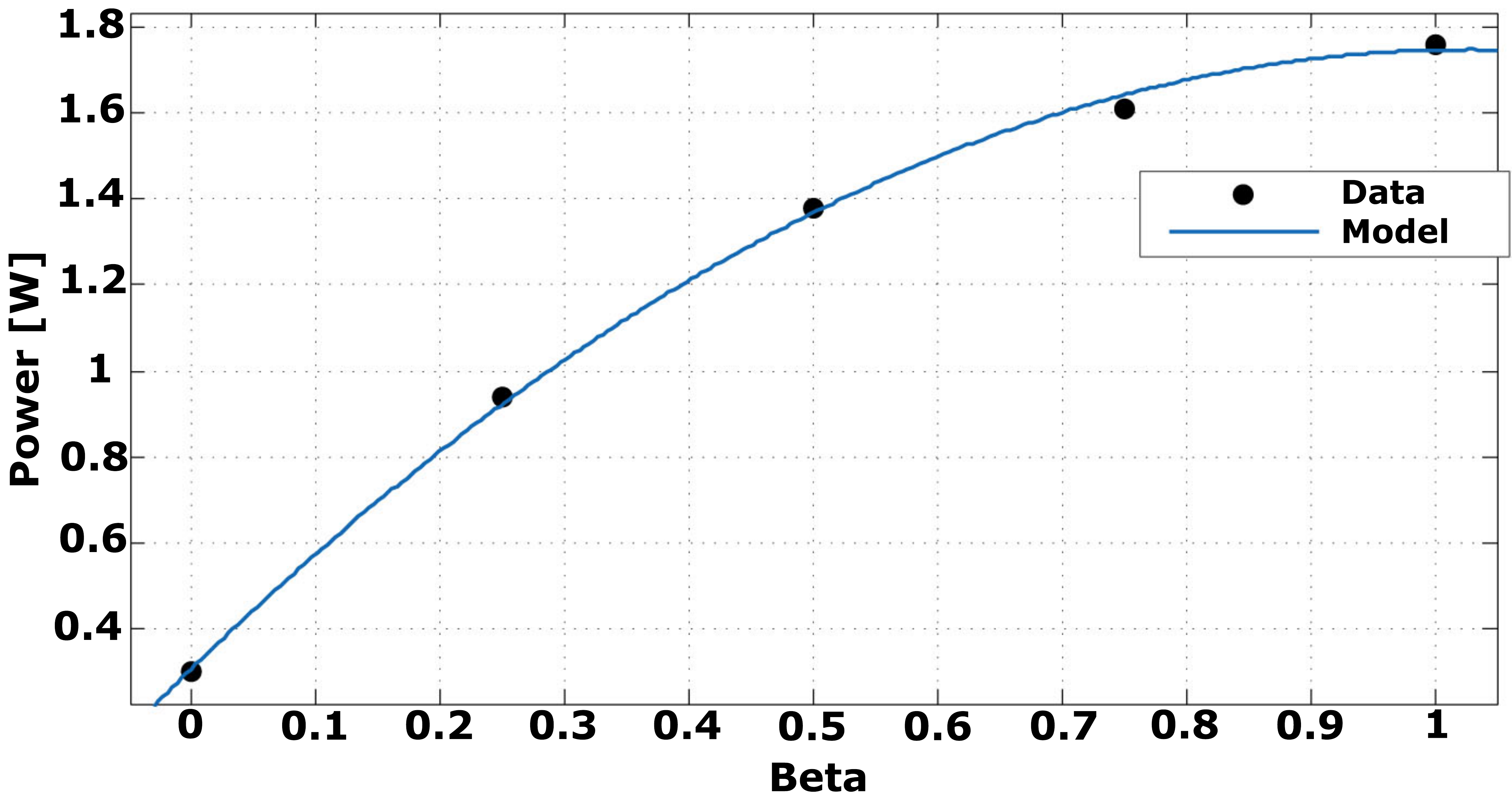}
\caption{Measurements of power dissipation based on I/O expressed as a ratio $\beta$. The model is expressed as a second degree polynomial.}
\label{fig:gammamodel}
\end{figure}

The explored I/O ratios ($\beta$ values) were selected in the range [0.0, 0.25, 0.5, 0.75, 1.0] where 0.0 indicates total I/O blocking and 1.0 indicates no I/O (a fully CPU-intensive workload).
The CPU (in this case the ARM architecture) was executing at 1600 MHz for all measurements (we show the behaviour of the model at different CPU frequencies in the following experiment),
and the measurement results are shown in Fig.~\ref{fig:gammamodel} as the data values.

A one-dimensional curve fitting technique was used to model the power ratio $\gamma_{core}$ as a second degree polynomial:
 \begin{align}
 \gamma_{core} = \frac{p_o \beta^2 + p_1 \beta + p_2}{P_{max}}
 \end{align}
where $\beta$ is the CPU-boundedness of the core, $P_{max}$ is the maximum power dissipation of a core and the obtained function parameters are listed in Table~\ref{tab:gammaparameters}.
The curve in Fig.~\ref{fig:gammamodel} shows the model of the $\gamma_{core}$ function.

\begin{table}[!b]
 \caption{\small Coefficients for the power ratio $\gamma_{core}$ model.}
  \begin{center}
  \begin{tabular}{ | c | c | c |}
  \hline
  $p_{0}$  & $p_{1}$ & $p_{2}$ \\ \hline
  \footnotesize -1.362   &\footnotesize 2.798  &\footnotesize 1.31  \\ \hline
  \end{tabular}
\label{tab:gammaparameters}
\end{center}
\end{table}

\begin{figure}[b]
\centering
\includegraphics[width=1.0\columnwidth]{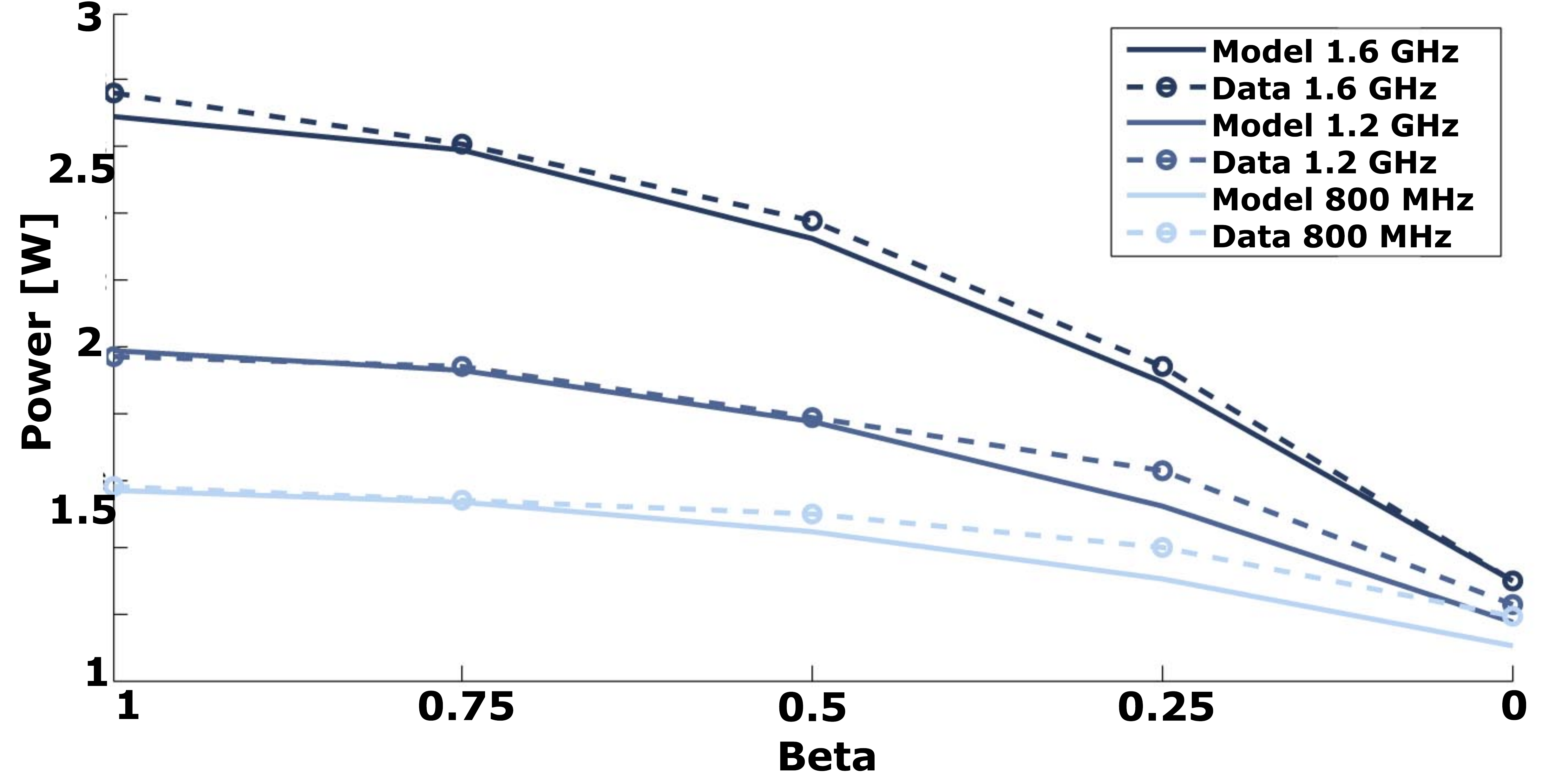}
\caption{Verification of the I/O-based power model from Fig.~\ref{fig:gammamodel}. Verification performed with three different frequency levels.}
\label{fig:gammaverify}
\end{figure}

The accuracy of the $\gamma_{cores}$ function at different CPU frequencies was evaluated in another experiment. We executed the Berserk benchmark with the same $\beta$ parameters at clock frequencies 1600MHz, 1200MHz and 800MHz.
Both the measurement data and the model for each experiment is shown in Fig.~\ref{fig:gammaverify}, in which the circles are measurement points.
The maximum difference between the data and the model was 10.59\% and the average difference was 4.32\%, which we considered as acceptable for our cloud controller evaluation.

To combine all of the presented model components, the power dissipation $P$ of a \gls{pm} in the cloud system was modelled as:
\begin{align}\label{eq:putil}
P = P_{idle} + (P_f - P_{idle}) u
\end{align}
where $P_{idle}$ is the idle power expressed in Eq.~\ref{eq:pfixed},
$P_{f}$ is the active power expressed in Eq.~\ref{eq:ppeak}
and $u$ is the utilisation modelled in Eq.~\ref{eq:utilisation}
based on the I/O activity parameter $\beta$. The model to compute the power dissipation of a \gls{pm} in the cloud system was based on the approach proposed in \cite{liu_renewable_2012}, where \gls{pm}'s power dissipation is linearly related with its CPU utilisation. The model was extended to take into account the active power dissipation of the currently used cores of the \gls{pm} at the operating frequency and the CPU-utilisation of the cores based on the \gls{vm} CPU-boundedness $\beta$.

\subsection{Energy Calculation}
\label{sec:energy_calculation}

With the multi-core, frequency-dependent power model, we now
present the details of electricity cost calculation based on geotemporal inputs.\
This conversion of power dissipation to a monetary value is crucial for comparing potential energy savings with revenue losses under performance-based \gls{vm} pricing that we explore in the next section.

The power model so far was expressed for instantaneous values,
but to express the energy costs for the cloud provider we need to add a time dimension
to account for geotemporal inputs and actions like frequency scaling.
Therefore, we time-stamp the expressions that change
as time progresses, so for example $P_t$ is the power dissipation for a \gls{pm} at time $t$
(as it depends on the current CPU frequency $f$).
We define an observed time period\
of $N$ equidistant time stamps\
in the range from $t_{0}$ to $t_{N}$, denoted $[t_{0}, t_{N}]$.\

Cooling overhead based on local temperatures\ 
is derived from the power signal of each \gls{pm} at its corresponding data center location.\
To do so, the model for computer room air conditioning using outside air economisers\
from \cite{xu_temperature_2013} was applied.\
Cooling efficiency is expressed as partial \gls{pue} -- $pPUE_{dc,t}$\
at data center $dc$ at time $t$,\
which affects the power overhead based on the following formula:
\begin{align}\label{eq:P_tot}
P_{tot,t} = P_t + P_{cool,t} = pPUE_{dc,t} \cdot P_t(pm)
\end{align}
where $P_{cool,t}$ is the power necessary to cool the physical machine, and\
$P_{tot,t}$ stands for the combined cooling and computation power.\
For each data center location $dc$, there is a time series of temperature values\
$\{T_{dc,t}:\ t \in [t_{0}, t_{N}]\}$.\
The dynamic value of $pPUE_{dc,t}$ is modelled as a function of\
temperature $T$ to match hardware specifics as:\
\begin{align}\label{eq:pPUE}
pPUE_{dc,t} = 7.1705 \cdot 10^{-5}T_{dc,t}^2 + 0.0041T_{dc,t} + 1.0743
\end{align}

These power signals are then integrated over time and combined with fixed
or real-time electricity prices (both models are explored in the evaluation)
for the corresponding data center location to compute the total energy cost $C_{en}$
of every individual \gls{pm}.
\vspace{\eqtopmargin}
\begin{align}\label{eq:price}
C_{en} = \frac{t_N-t_0}{N} \mathlarger{\sum}_{t=t_0}^{t_N-1} P_{tot,t}e_{dc,t}
\end{align}
The integration is approximated using the rectangle numerical integration method.\
At each $dc$ location, there is a time series of electricity prices\
$\{e_{dc,t}:\ t \in [t_{0}, t_{N}]\}$.\




\section{Performance-Based \gls{vm} Pricing}
\label{sec:pricing_model}

After covering the detailed components that make up energy costs in modern multi-core physical machines, in this section we continue analysing the economical side of cloud computing by looking at \gls{vm} pricing. Concretely, we cover state of the art performance-based \gls{vm} pricing schemes used by providers such as ElasticHosts and CloudSigma where the user pays for the \gls{vm} proportionally to the allocated CPU frequency. We then show on a practical experiment that these schemes do not account for properties like the CPU-boundedness of the \gls{vm}'s workload and its effect on \gls{qos} in the price calculation. To address such drawbacks, we present our own perceived-performance \gls{vm} pricing scheme as a next step in performance-based pricing, adapted for both Intel and ARM architectures. Having models for both the energy costs and \gls{vm} revenue accounting for frequency scaling on multi-core \gls{pm}s will allow us to explain our cloud controller in the next section. 



\subsection{Emerging Performance-Based Pricing \\Cloud Providers}

In the performance-based pricing model used by several cloud providers, each user is charged on a per-time-unit basis (e.g. hourly) depending on the provisioned resources and their characteristics. The overall cost includes the cost for CPU provisioning, the allocated amount of RAM and the use of other resources, e.g. storage. Such a pricing scheme is offered by several providers, such as ElasticHosts \cite{elastichosts} and CloudSigma \cite{cloudsigma}, that allow the provisioning of different CPU frequency and core quantities, calculating the total CPU capacity allocated for the final invoice. This enables users to choose between equivalent combinations of frequencies and number of virtual CPU cores that incur same CPU provisioning costs \cite{elastichosts}. 

Based on the performance-based pricing scheme offered by ElasticHosts and CloudSigma, we fitted a pricing model that describes the behaviour of both schemes, similarly to the work in \cite{pietri2015cost} where the ElasticHosts pricing scheme was initially modelled and analysed. In our obtained model, the price charged for CPU provisioning changes linearly with the total requested CPU capacity, as CPU capacity can be customised for different selected CPU frequencies and number of cores. Also, we extended the model to describe the RAM allocation. As \gls{vm}s may have different RAM capacity requirements, the price varies according to the selected RAM size. Finally, the cost for other resources used is considered to be fixed in the model as it is not the focus in this work. 
Hence, the price $C_{vm}$ of each \gls{vm} at frequency $f_{CPU}$ is computed as:
\begin{align}\label{eq:simpleprice}
C_{vm} = C_{base} + C_{CPU} \sum \limits_{cpu \in vm} {(\frac{f_{cpu} - f_{base}}{f_{base}})} + C_{RAM} \frac{RAMsize}{RAMsize_{base}},
\end{align}
where $C_{base}$ is the price of the \gls{vm} at minimum capacity, i.e. a CPU at minimum frequency $f_{base}$. $C_{CPU}$ and $C_{RAM}$ are cost weights used to generate the \gls{vm} price for different CPU and RAM capacities. By replacing these variables with actual constants (presented later in the evaluation), the prices for configurations offered by ElasticHosts or CloudSigma can be approximated.


\subsection{Workload Heterogeneity Implications}

While the performance-based pricing model offered by ElasticHosts and CloudSigma enables the cloud provider to balance energy savings with the revenue losses from actions such as CPU frequency scaling, it ignores the impact of \gls{vm} workload characteristics.\
We illustrate this in an empirical experiment we have performed to show how operating CPU frequency may affect workload performance in a different way depending on the application's CPU boundedness ($\beta$) characteristics.\

We executed the Berserk benchmark (already explained in Section~\ref{section:berserk}) on one local server with a \emph{remote\_ratio} parameter indicating the portion of the work to offload to a different, remote server. The rest of the tasks were executed locally. Both servers had the same Quad-core 2.6 GHz AMD Opteron 4130 processor. The \emph{remote\_ratio} parameter therefore controlled the workload's CPU boundedness, as we could control if the task was more CPU-bound (i.e. a low \emph{remote\_ratio}) or more I/O-bound where they would wait on the results to arrive from a network resource (i.e. a high \emph{remote\_ratio}). We calculated the CPU boundedness parameter $\beta$ as inversly proportional to \emph{remote\_ratio}.\
This approach enabled us to set arbitrary workload CPU boundedness. The experiment was run for six equidistant $\beta$ values between 0 and 1. 

To also measure the effects CPU frequency scaling has in this context, we executed each of the workload characteristics on all five CPU frequency levels (applied both locally and to the remote server) that our server offered using the `cpufrequtils` tool (the scripts we developed are included together with the Berserk benchmark).
We collected the duration of the benchmark under each of the workload CPU boundedness $\beta$ and server CPU frequency combination.\


 \begin{figure}
 \centering
 \includegraphics[trim=2.5cm 1cm 1.2cm 1.5cm, clip=true, width=0.95\columnwidth]{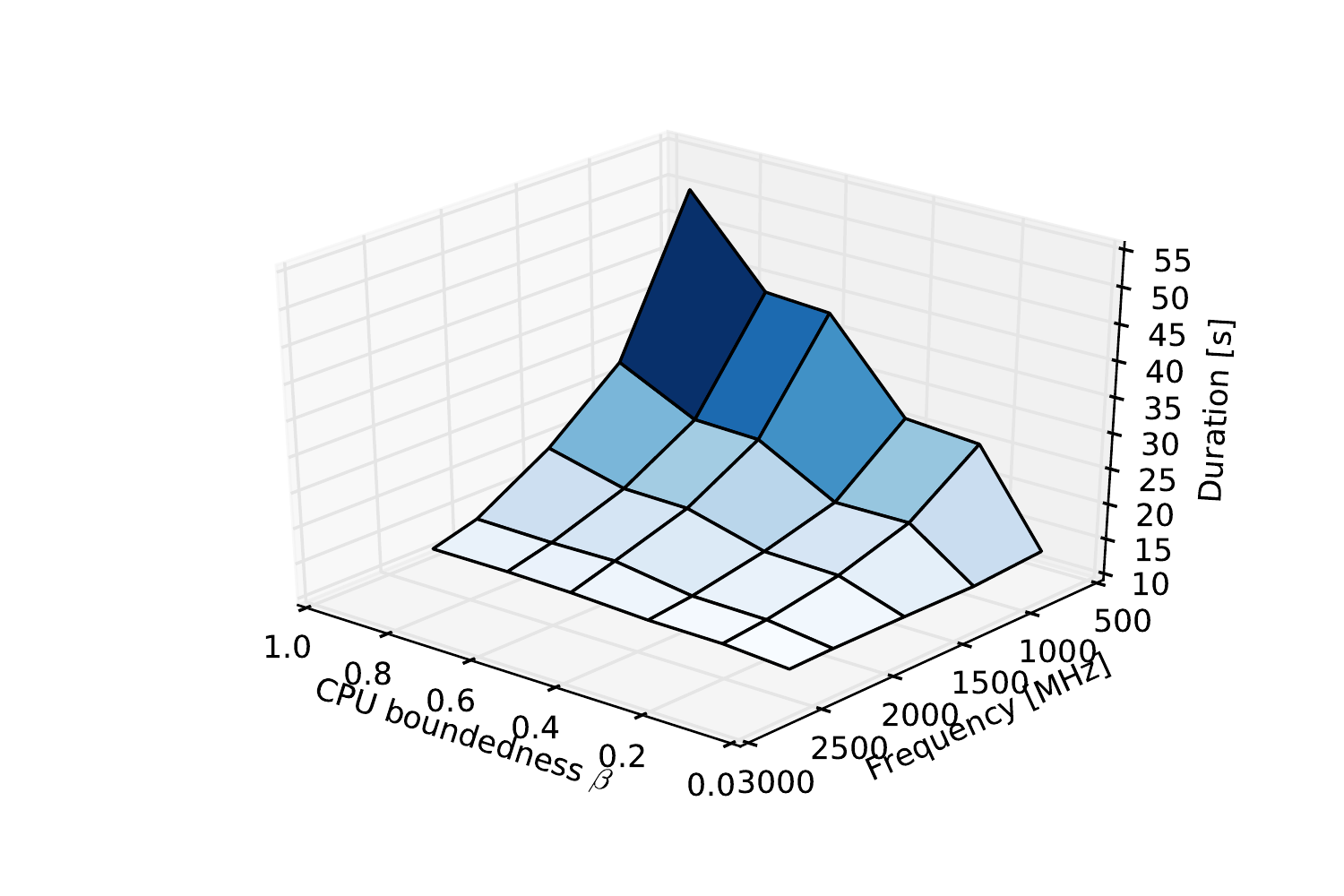}
 \caption{Benchmark duration for different workload CPU-boundedness ($\beta$) parameters and server CPU frequencies.}
 \label{fig:motExample}
 \end{figure}



The results are shown in Fig.~\ref{fig:motExample}. As can be seen, the execution time of an application with high CPU-boundedness ($\beta$) increases significantly when using a lower frequency and remains mostly unaffected for application with a lower CPU-boundedness ($\beta$ close to 0). Using the current performance-based pricing for CPU provisioning, like ElasticHosts and CloudSigma, a low frequency for jobs with low CPU boundedness would result in significantly lower revenue for the provider, even though the application performance would not be greatly affected. This was the main motivation for our perceived-performance pricing scheme that we present in the following section.


\subsection{Perceived-Performance Pricing} \label{section:perc_pricing}

 \begin{figure}[b]
 \centering
 \includegraphics[width=0.9\columnwidth]{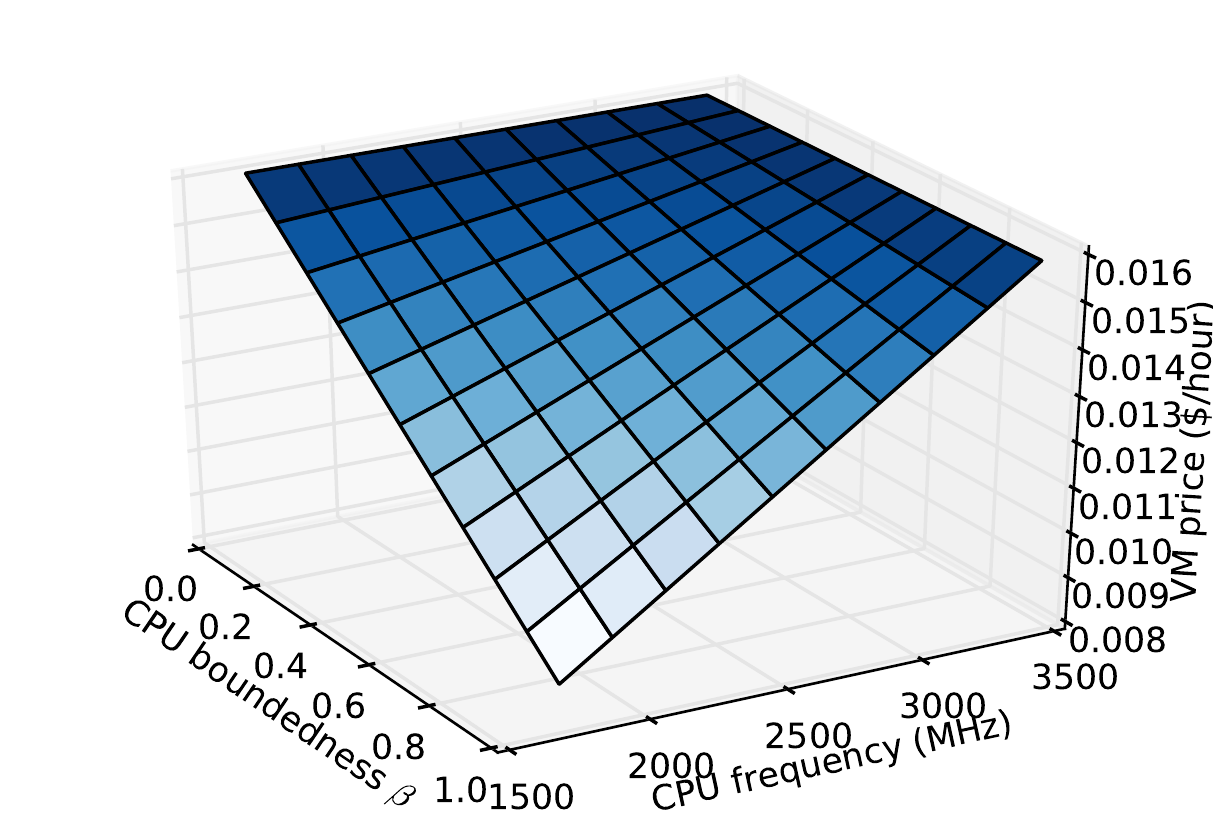}
 \caption{Perceived-performance pricing model.}
 \label{fig:pricing_model}
 \end{figure}
 
As mentioned earlier, performance-based pricing, used by cloud providers like ElasticHosts and CloudSigma, does not consider the impact of frequency reduction on \gls{vm} performance. To do so, Eq.~\ref{eq:simpleprice} is extended for pricing based on the performance perceived by the hosted application based on the approach we proposed in \cite{lucanin_cloud_2015} by defining $f_{cpu}$ as:
\
\begin{align}\label{eq:perc_pricing}
f_{cpu} = \beta f + (1 - \beta) f_{max},
\end{align}
where $f_{max}$ is the maximum operating frequency of the host 
so that CPU-bound applications incur lower CPU provisioning cost when using lower frequencies that may result in lower performance. On the other hand, I/O-bound applications that are not affected by the change in frequency do not receive significant decrease in cost. It is assumed that the impact of frequency on application performance is same for each core of the \gls{vm} ($\beta$).

The pricing model is presented in Fig.~\ref{fig:pricing_model}, where the axes show the provisioned CPU capacity, the CPU boundedness of the \gls{vm} (parameter $\beta$) and the respective \gls{vm} price. The CloudSigma model constants were used. The amount of RAM was not varied. CPU frequency values for the Intel platform were used. We assumed four active CPU cores where the CPU frequency is scaled linearly (so the sum of assigned CPU frequencies of all the cores was used in the price calculation).

The linear curve for $\beta=1$ corresponds to the case of performance-based pricing where the CPU provisioning cost changes linearly with the actual CPU frequency ($f_{cpu} = f$). As the value of $\beta$ is decreased, the price is less affected by CPU scaling (becoming constant for $\beta=0$), which matches the workload behaviour from the experiment described earlier.

It is assumed that resources are charged on an hourly basis. The total service revenue is computed by adding the per-hour provisioning cost of Eq.~\ref{eq:simpleprice} for each \gls{vm} served, allowing us to compare it with the frequency-based energy costs from Section~\ref{sec:power_model}.

\subsection{Prices for Different Architectures}
\label{sec:pricing_model_arm}

As mentioned in Section~\ref{sec:mont_blanc}, besides using Intel architectures with ElasticHosts and CloudSigma pricing, in our work we are also interested in analysing the new pricing models on ARM architectures. Since the performance of an ARM-based CPU is significantly lower than the Intel, the price was scaled according to this factor.
The authors in \cite{chi2015benchmark} evaluated the ARM Cortex-A15 against a Haswell i5 executing \gls{hevc} decoding tasks and obtained a 6x performance advantage for the Intel.
Similarly, the authors in \cite{francesquini2015benchmark} evaluated a Sandy-Bridge based Xeon and a Cortex-A15 using various benchmarks, with roughly an 8x performance advantage for the Intel.
Finally, a set of scientific benchmarks was used in \cite{padoin2015benchmark} with both a Cortex-A15 CPU and a Sandy-Bridge based high-end i7 CPU.
The results indicated that the i7 performed roughly 16x better than the ARM.

Based on these numbers, and by using one ARM Cortex-A15 CPU and one Sandy-Bridge based high-end i7,
we normalised the performance values to the clock frequency of both platforms in order to match with the cost model of the cloud provider.
We assumed a 11x performance advantage for the Intel architecture, and therefore we assumed a 11x cost reduction in the cloud service when using the ARM platform.


\section{Cloud Controller Description}
\label{sec:scheduler}

Having described both the multi-core, geographically-dependent energy consumption model used to compute the energy costs from operating the cloud and the perceived-performance pricing scheme used to compute the revenue from \gls{vm} provisioning, in this section we explain our cloud controller. The cloud controller balances both of these cost-related components\
to obtain a quantitative comparison of energy saving and revenue loss trade-offs,
which can be addressed as an optimisation problem.\ 
In other words, both cost-related components addressed in this paper are used to determine the actions invoked by the cloud controller. 

We describe the \schedulerfreq{} cloud controller that determines the \gls{vm} migration and frequency scaling actions to be triggered in order to achieve energy cost savings that exceed the revenue losses. As these two control actions -- \gls{vm} migration and CPU frequency scaling -- are mutually orthogonal, they are considered as two complementary actions in order to optimise the allocation and configuration of \gls{vm}s to \gls{pm}s. Hence, the two actions are examined in the algorithm separately as two stages, firstly migrating the \gls{vm}s to \gls{pm}s and then scaling the CPU frequencies of the \gls{pm}s to achieve further energy cost savings. During the \gls{vm} migration stage, the controller migrates \gls{vm}s to \gls{pm}s so that resource utilisation is maximised while preferring more economical locations in terms of electricity and cooling costs. Then, the controller reduces the CPU frequencies of the \gls{pm}s iteratively as long as the energy cost savings exceed the service revenue losses. The algorithm is invoked periodically to trigger appropriate actions. In a real cloud system the algorithm could be automatically invoked by new \gls{vm} arrivals or threshold violations of geotemporal inputs, e.g. a temperature increase of 1 C. Next, the two stages of the algorithm are described in more detail.


\subsection{\gls{vm} Migration Stage}

\begin{algorithm}[!t]
\caption{\gls{vm} Migration Stage.}
\label{alg:bcf}
{
\begin{algorithmic}[1]

\Ensure Allocate or migrate \gls{vm}s per geotemporal inputs.
\Procedure{\gls{vm} Migration Stage}{}
\State $to\_alloc \gets$ empty list
\State append all \gls{vm}s newly requested to $to\_alloc$ \label{alg:bcf:boot} 
\State append \gls{vm}s from all underutilised \gls{pm}s to $to\_alloc$ \label{alg:bcf:underutil}
\State sort $to\_alloc$ by resource requirements decreasing \label{alg:bcf:vmsort}
\For{$vm \in to\_alloc$} \label{alg:bcf:vmloop}
    \State $active \gets $ all PMs where at least one VM is allocated 
    \State $inactive \gets $ all PMs where no VMs are allocated
    \State sort $inactive$ by capacity decreasing, cost increasing\
    \label{alg:bcf:sortinactive}
    \State $mapped \gets False$
    \While{not $mapped$} \label{alg:bcf:mappedloop}
        \State sort $active$ by capacity decreasing, cost increasing \label{alg:bcf:sortactive}
        \For {$pm \in active$}\label{alg:bcf:pmloop}
            \If {$vm$ fits $pm$}\label{alg:bcf:pmfits}
                \State $mapped \gets True$
                \State break loop
            \EndIf
        \EndFor
        \If {not $mapped$}
            \State pop $inactive[0]$ and append it to $active$ \label{alg:bcf:popinactive}
        \EndIf
    \EndWhile
    \State perform a placement/migration of $vm$ to $pm$ \label{alg:bcf:execute}
\EndFor
\EndProcedure

\end{algorithmic}
}
\end{algorithm}

During the first stage, the controller allocates newly requested \gls{vm}s or reallocates \gls{vm}s from underutilised hosts using migration based on the power overhead and the geotemporal input parameters of the \gls{pm}s. As the underlying bin packing problem of \gls{vm} allocation is NP-hard, we propose a heuristic polynomial time algorithm.

The \gls{vm} migration stage pseudo-code is shown in Alg.~\ref{alg:bcf}. The algorithm initially marks for allocation all the newly requested \gls{vm}s (line~\ref{alg:bcf:boot}) and for reallocation all \gls{vm}s that run on underutilised hosts (line~\ref{alg:bcf:underutil}), considering hosts as underutilised if their utilisation falls below a provider-defined threshold, as discussed in \cite{beloglazov_managing_2013}.\
The selected  \gls{vm}s (line~\ref{alg:bcf:vmloop}) are then migrated (or initially placed), prioritising \gls{vm}s larger in their resource requirements (e.g. more required RAM, CPU cores), which are more difficult to fit (line~\ref{alg:bcf:vmsort}). Then, the available \gls{pm}s are divided into $active$ and $nonactive$ lists depending on their state (suspended or not). \gls{pm}s in the $inactive$ list are sorted (line~\ref{alg:bcf:sortinactive}) so that larger \gls{pm}s are preferred to smaller machines (in order to minimise the idle power overhead) and data centers with lower combined electricity price and cooling overhead cost are prioritised based on the geotemporal input prices model presented in Section~\ref{sec:energy_calculation}.\
The \gls{pm} that will act as a $vm$ host is selected by sorting $active$ so that almost full \gls{pm}s are utilised first, preferring \gls{pm}s at lower-cost locations in case of ties (line~\ref{alg:bcf:sortactive}). When the $vm$ does not fit on any of the $active$ \gls{pm}s, the next \gls{pm} from $inactive$ is activated (line~\ref{alg:bcf:popinactive}). Again, \gls{pm} sorting assures that data centers will be selected based on the current geotemporal inputs (Section~\ref{sec:energy_calculation}). When a host \gls{pm} is found, the \gls{vm} is placed or migrated to it (line~\ref{alg:bcf:execute}) and the algorithm continues with the next \gls{vm}.

\subsection{Frequency Scaling Stage}

\begin{algorithm}[!t]
\caption{Frequency Scaling Stage.}
\label{alg:frequency_scaling}
{
\begin{algorithmic}[1]

\Ensure Reduce CPU frequencies while energy savings exceed revenue losses.
\Procedure{Frequency Scaling Stage}{}

\State $decrease\_feasible \gets False$
\State reset frequency of $\forall pm \in active$ to $f_{max}$\
    \label{alg:fs:reset}

\For {$pm \in active$} \label{alg:fs:pmiterate} 
    \State $f \gets f_{max}$ \
        \Comment{Start the loop at max frequency}
    \State $revenue\_cur \gets get\_revenue(pm, f_{to\_apply})$ 
    
    \Comment Service revenue, $\forall vm \in pm$
    \State $en\_cost\_cur \gets get\_en\_cost(pm, f_{to\_apply})$ \\ \Comment Energy cost of the $pm$

    \While{$f>f_{min}$} 

        \State $f \gets f-f_{step}$ \
            \label{alg:fs:freqStep}
        \State $revenue\_new \gets get\_revenue(pm, f)$ \
        
            \Comment {Revenue for the new frequency} \
            \label{alg:fs:get_revenue}
        \State $en\_cost\_new \gets get\_en\_cost(pm, f)$ \\
            \Comment {New energy cost} \
            \label{alg:fs:get_cost}
        \State $revenue\_loss \gets revenue\_cur - revenue\_new$ \
        \State $en\_savings \gets en\_cost\_cur - en\_cost\_new$ \
        \If {$en\_savings > revenue\_loss$} \
            \label{alg:fs:comparison}
            \State $revenue\_cur \gets revenue\_new$ \\
            
                \Comment{Update current service revenue}
            \State $en\_cost\_cur \gets en\_cost\_new$ \\
                \Comment{Update current energy cost}
            \State $decrease\_feasible \gets True$ \
            \State $f_{to\_apply} \gets f$ \\
                \Comment{Update currently selected frequency} \
                \label{alg:fs:apply}
        \Else
            \State break \label{alg:fs:break}
        \EndIf
    \EndWhile

    \If {$decrease\_feasible$} \
        \State apply $f_{to\_apply}$ to $pm$
    \Else
        \State remove from $active$: $\forall \hat{pm} \in PMs$ s.t. $\hat{pm}$ has higher mean $\beta_{vm})$ and lower el. price and temperature than $pm$ \
        \label{alg:fs:remove_worse}
    \EndIf
\EndFor
\EndProcedure

\end{algorithmic}
}
\end{algorithm}

\begin{figure}[!b]
\centering
\includegraphics[width=1.0\columnwidth]{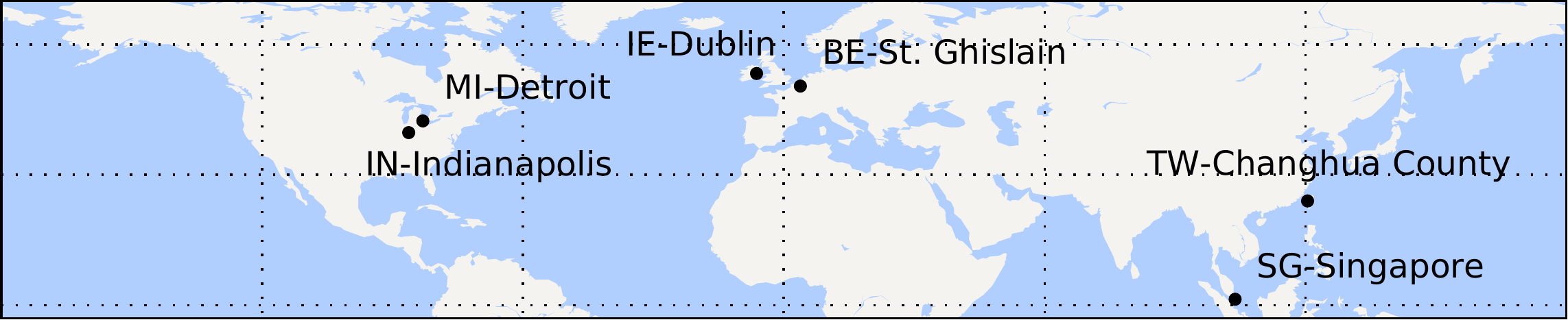}
\caption{Cities used as data center locations in the simulation.}
\label{fig:cities}
\end{figure}

Having allocated the \gls{vm}s to \gls{pm}s, the CPU frequencies of the \gls{pm}s are adjusted in the next stage. We assume that each host can operate between a minimum and maximum frequency, $f_{min}$ and $f_{max}$ respectively. The appropriate CPU frequencies are selected based on both the geotemporal inputs and the workload characteristics, by considering their overall impact on the cost components presented in Sections~\ref{sec:power_model} and~\ref{sec:pricing_model}. To do so, a \gls{pm}'s CPU frequency is reduced only when energy savings from the reduction in the CPU frequency exceed the revenue losses under perceived-performance pricing. The algorithm is described in Alg.~\ref{alg:frequency_scaling}. From a high level, the CPU frequency of each \gls{pm} is initially set to its maximum frequency $f_{max}$ (line~\ref{alg:fs:reset}). Then, the algorithm iterates through the list of $active$ \gls{pm}s (line~\ref{alg:fs:pmiterate}) to determine the most efficient CPU frequency for each one, analysing the range of the available CPU frequencies (line~\ref{alg:fs:freqStep}). Note that the actions determined in each step do not have to be executed physically before the procedure halts where the final \gls{pm} frequencies are determined.

To pick the best CPU frequency, the cost-related components for the current \gls{pm} are calculated for the previously determined and the next lower frequency (lines~\ref{alg:fs:get_revenue}--\ref{alg:fs:get_cost}). The components include the service revenue from the \gls{vm}s allocated to the current \gls{pm} and the \gls{pm}'s energy cost based on the multi-core power model and energy cost calculation presented in Section~\ref{sec:power_model}. Whenever the consideration of the lower CPU frequency results in energy cost savings which exceed the subsequent revenue loss, the new frequency is chosen for the current \gls{pm} (line~\ref{alg:fs:comparison}) and the algorithm continues to the next lower available frequency (line~\ref{alg:fs:freqStep}). The procedure in the inner loop terminates when the revenue losses exceed the respective energy savings (line~\ref{alg:fs:break}). If no frequency reduction occurred for the \gls{pm} ($decrease\_feasible$ stays $False$), the procedure will remove \gls{pm}s with higher average $\beta$ and lower electricity price and temperature (line~\ref{alg:fs:remove_worse}) before continuing. The idea is that such \gls{pm}s may incur even lower energy savings and higher revenue losses, hence they can be omitted from the analysis to prune the search space. 


\section{Evaluation}
\label{sec:evaluation}

In this section we evaluate the presented cloud controller in a simulation that we first describe as part of the evaluation methodology. We then proceed with presenting the simulation results showing the impact of our cloud controller with a focus on different environment factors.

\subsection{Methodology}

The \schedulerfreq{} method is evaluated in a\
simulation of \vmnumsimulation{} \gls{vm}s based on real traces of geotemporal inputs and \gls{vm} CPU-boundedness values. The goal of the evaluation is to show the cost savings attainable using our approach, the impact on service revenue and to analyse the dependence on external factors, such as electricity prices and \gls{vm} workloads. 
The simulations were executed on our open source Philharmonic\ 
simulator framework~\cite{drazen_lucanin_philharmonic_2014}. A simulation in Philharmonic consists of iterating through the timeline, collecting the currently available electricity prices and temperatures, as well as the incoming \gls{vm} requests. The simulated controller is called to determine cloud control actions, such as \gls{vm} migrations or \gls{pm} frequency scaling. The applied actions are used to compute the resulting energy consumption and electricity costs based on our cost model from Section~\ref{sec:power_model}.

To compute the energy costs of the simulated geographically-distributed cloud, we consider a use case of six data centers. A dataset of real-time electricity prices described in \cite{alfeld_toward_2012}\
and temperatures from the Forecast~\cite{_forecast_2015} 
web service were used. The data center locations used in the simulation (Fig.~\ref{fig:cities}) were selected to resemble Google's deployment. Due to lack of \gls{rtep} data for the four non-US cities, the electricity prices were synthetically generated from the data known for other US cities -- the time series were shifted based on the time zone offsets and a difference in annual mean values was added in order to resemble local values. Additionally, a scenario with fixed electricity prices over time is considered in the evaluation, using the mean values for each location.

\begin{table}[!t]
\centering
\caption{Infrastructure parameters.}
\label{tab:simulation}
\begin{tabular}{cccccc}
\hline
 Architecture & \gls{pm}s & \gls{vm}s & \
 $f_{min}$ & $f_{max}$ & $f_{step}$ \\
\hline
 ARM & \pmnumsimulation{} & \vmnumsimulation{}& \
 0.8 GHz & 1.8 GHz & 100 MHz \\
\hline
 Intel & \pmnumsimulation{} & \vmnumsimulation{}& \
 2.6 GHz & 3.4 GHz & 200 MHz \\
\hline
\end{tabular}
\end{table}

\begin{table}[!h]
\centering
\caption{Pricing model parameters.}
\label{tab:pricing_simulation}
\begin{tabular}{ccccccccccccc}
\hline
 Pricing Model & $C_{base}$ & $C_{CPU}$ & $C_{RAM}$ &\
 $RAMsize_{base}$  \\
\hline
ElasticHosts & 0.027 \$/h  & 0.018 \$/h  & 0.025 \$/h & 1 GB\\
\hline
CloudSigma & 0.0045 \$/h  & 0.0017 \$/h  & 0.004 \$/h & 1 GB\\
\hline
\end{tabular}
\end{table}

The simulator was set up using the infrastructure parameters shown in Table~\ref{tab:simulation}.\
The table shows two architecture types: ARM and Intel. Their respective performance characteristics were derived from the real specifications, such as minimum CPU frequency $f_{min}$, maximum CPU frequency $f_{max}$ and the absolute frequency increase or decrease step size $f_{step}$.\
\
\
The parameters we fitted for the pricing models in Section~\ref{sec:pricing_model} to calculate hourly \gls{vm} prices based on the pricing schemes offered by ElasticHosts \cite{elastichosts} and CloudSigma \cite{cloudsigma} are shown in Table~\ref{tab:pricing_simulation}. The cost of other resources which is not the focus in this work, e.g. disk, was considered to be fixed.
\
Due to space restrictions, we show results for both CPU architectures and both pricing schemes only in sections where we compare the effects of these respective factors on the attainable energy and cost savings. Other presented results are limited to the ARM architecture and the CloudSigma pricing scheme, which proved to be more promising for the application of our method, as will be shown later.\

Each run simulated the cloud system for seven days of operation (168 h) with an hourly step size (1 h). The step size was chosen based on the available datasets of geotemporal inputs. However, note that different time intervals and triggering events, e.g. thresholds in geotemporal input changes or new \gls{vm} arrivals could invoke the cloud controller in production environments.\
The characteristics of each resource considered in this work, namely the number of CPU cores and the amount of RAM, were uniformly distributed.\
Heterogeneous \gls{vm}s were assumed with 1 or 2 CPU cores and RAM capacity ranging between 8 and 16 GB RAM in order to model different \gls{vm} requests and prices. Each \gls{pm} consists of 1--4 CPU cores and 16--32 GB RAM to model specification diversity. For each \gls{vm}, the boot time and duration were varied using a uniform distribution to generate random values within the simulation time and distribute delete events over the simulation period and range the utilisation of the resources.

\begin{figure}[!b]
\centering
\includegraphics[width=0.85\columnwidth]{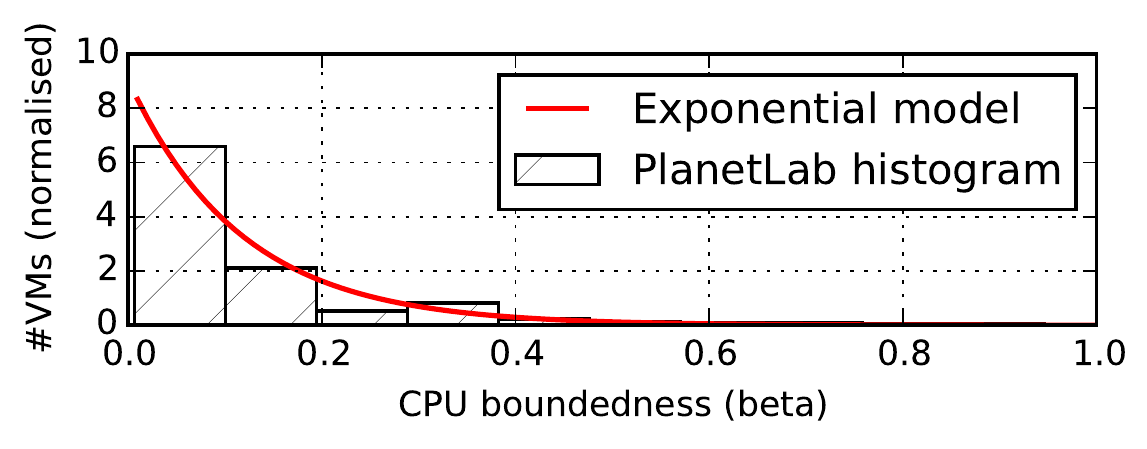}
\caption{\gls{vm} CPU-boundedness distribution from PlanetLab traces.} 
\label{fig:beta_model}
\end{figure}

The CPU-boundedness of each \gls{vm} was modelled based on the CPU usage traces from the PlanetLab dataset \cite{planetlab}.\
The dataset includes CPU usage traces of 1024 \gls{vm}s. The data was collected every five minutes throughout a day. To generate realistic \gls{vm} CPU-boundedness values in the simulation, the average CPU usage of each \gls{vm} in the dataset was calculated and mapped to a $\beta$ value. From the generated dataset of $\beta$ values, an exponential distribution was fitted. The distribution is shown in Fig.~\ref{fig:beta_model}. The figure also includes the empirical histogram of the traces normalised to an area of 1. The $\beta$ values of the \gls{vm}s used in the simulation were generated based on this model.




We consider two baseline controllers for results comparison.\
The \schedulerbase{} algorithm developed in \cite{beloglazov_energy-aware_2012} is a cloud controller that migrates \gls{vm}s, dynamically adapting to user requests. The second baseline controller, \schedulermigr{}, is a variant of the \schedulerfreq{} controller that applies \gls{vm} migration based on geotemporal inputs, but does not consider frequency scaling. The \schedulerfreq{} controller allows us\
to quantify the improvement brought by CPU frequency scaling in isolation. 

The remainder of the section presents individual results for the different simulation scenarios we performed to compare the energy and cost savings and the performance implications from applying the proposed cloud controller approach. The parameters specified earlier are used in all of the experiments, unless otherwise stated.


\begin{figure}[!t]
\vspace{\figtopmargin}
\centering
\includegraphics[width=1.0\columnwidth]{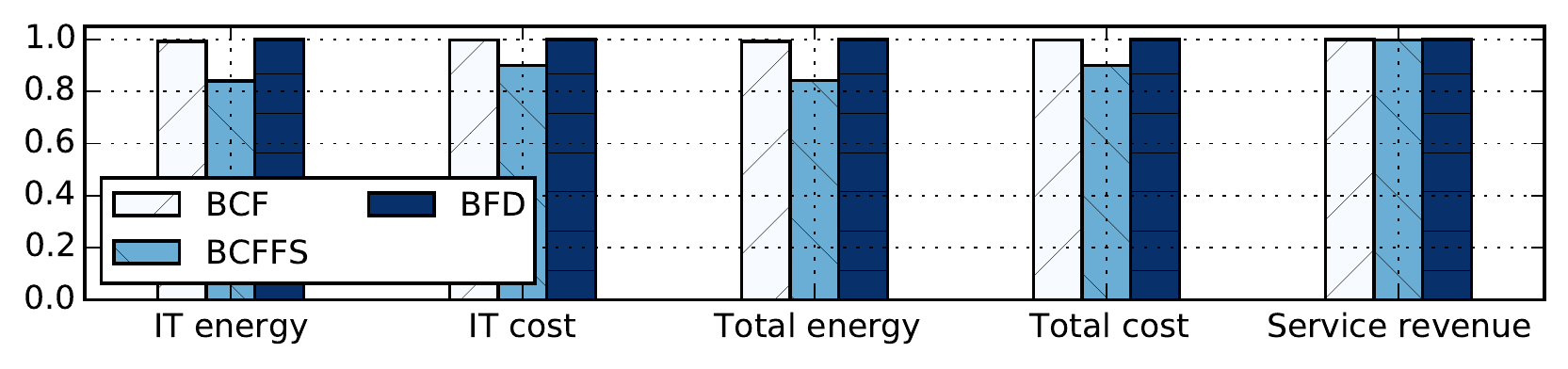}
\vspace{\figcaptionmargin}
\caption{Aggregated results for a \pmnumsimulation{} Intel PM simulation.} 
\label{fig:aggregated_results-single_simulation}
\vspace{\figbottommargin}
\end{figure}

\subsection{Cloud Controller Evaluation}

We begin by showing the cloud controller evaluation and comparison to the baselines for the Intel architecture. Fig.~\ref{fig:aggregated_results-single_simulation} includes the aggregated results for the achieved energy costs and service revenue \
-- the energy and cost used by the IT equipment, total energy and cost which include the cooling overhead taking into account the outside temperatures and the service revenue from hosting \gls{vm}s considering the perceived-performance pricing model described in Section~\ref{section:perc_pricing}. 
The values are normalised as a relative value of the results obtained using the baseline \schedulerbase{} controller, while the absolute values can be found in Table~\ref{table:aggregated_results-single_simulation}. It can be seen that the \schedulerfreq{} controller achieves \ensavingsmaxintel{} energy savings compared to the baseline controller \schedulerbase{}, out of which \ensavingsfreqintel{} are the additional energy savings achieved by using frequency scaling (the savings compared to the \schedulermigr{} controller). The service revenue losses from using perceived-performance pricing were not significant with a drop of less than 0.3\%, compared to the \schedulerbase{} baseline. This is because the frequency scaling algorithm presented in the previous section does not scale frequencies if the revenue loss exceeds the energy cost savings.



\begin{table}[!b]
\centering
\caption{Absolute aggregated Intel simulation results.}
\begin{tabular}{lrrr}
\toprule
{} &        BCF &      BCFFS &        BFD \\
\midrule
IT energy (kWh)     &   18793.73 &   15933.10 &   18943.00 \\
IT cost (\$)         &     974.60 &     878.50 &     977.00 \\
Total energy (kWh)  &   22501.25 &   19095.69 &   22678.65 \\
Total cost (\$)      &    1161.18 &    1046.75 &    1163.89 \\
Service revenue (\$) &    6543.78 &    6524.54 &    6543.78 \\
\bottomrule
\end{tabular}
\label{table:aggregated_results-single_simulation}
\end{table}


\subsection{Architecture Impact}

Having shown the results for the Intel architecture, we now show results for the same simulation, only this time using the ARM power model, presented in Section~\ref{sec:power_model_arm}. This allows us to compare the architecture impact on attainable energy cost savings. As we previously mentioned, ARM processors are increasingly popular due to their good computation-per-watt ratio and are being explored for use in data centers as part of the Mont Blanc project \cite{francesquini2015benchmark}. The \gls{vm} pricing is also adapted for the lower ARM performance compared to Intel, as detailed in Section~\ref{sec:pricing_model_arm}.

\begin{figure}[!t]
\centering
\includegraphics[width=1.0\columnwidth]{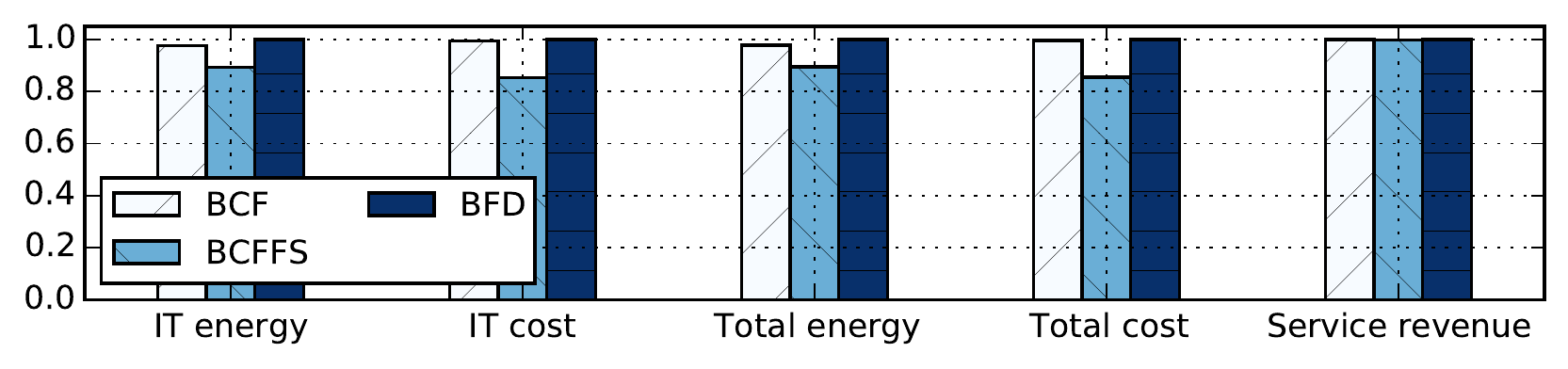}
\caption{Aggregated results for the ARM power and pricing model.}
\label{fig:aggregated_results-ARM}
\end{figure}

The results for the ARM architecture are shown in Fig.~\ref{fig:aggregated_results-ARM}. Even higher savings are achieved than for the Intel architecture -- the \schedulerfreq{} controller achieves \ensavingsmax{} energy cost savings compared to the \schedulerbase{} baseline and \ensavingsfreq{} compared to the \schedulermigr{} baseline. Even lower service revenue losses (less than 0.25\% drop compared to \schedulerbase{}) again indicate that the impact on \gls{vm} performance is not significant. Given the better applicability of our controller method to ARM architectures, we limit the remainder of the results to this architecture. 
Absolute values can be found in Table~\ref{table:aggregated_results-single_simulation-arm}.


\begin{table}[!b]
\centering
\caption{Absolute aggregated ARM simulation results.}
\begin{tabular}{lrrr}
\toprule
{} &        BCF &      BCFFS &        BFD \\
\midrule
IT energy (kWh)     &     681.79 &     623.50 &     698.19 \\
IT cost (\$)         &      39.23 &      33.68 &      39.48 \\
Total energy (kWh)  &     817.12 &     747.47 &     835.61 \\
Total cost (\$)      &      46.78 &      40.17 &      47.02 \\
Service revenue (\$) &     588.81 &     587.33 &     588.81 \\
\bottomrule
\end{tabular}
\label{table:aggregated_results-single_simulation-arm}
\end{table}


\subsection{Dynamic CPU frequency analysis}

\begin{figure}[!t]
\centering
\includegraphics[width=0.9\columnwidth]{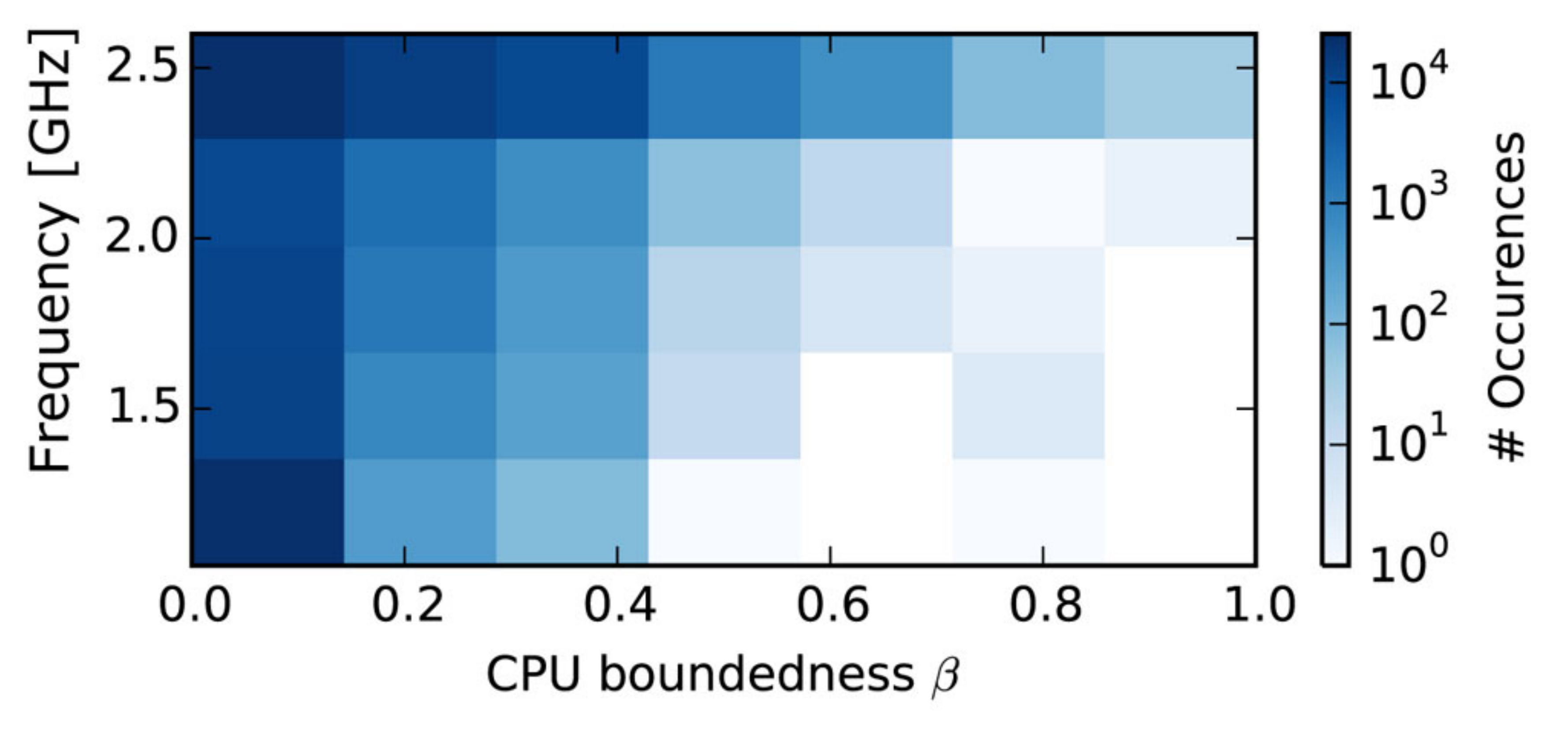}
\caption{Occurrences of $(\beta, f)$ combinations among the controlled \gls{vm}s for the ARM architecture.}
\label{fig:beta_freq_histogram}
\end{figure}

To explore the frequencies $f$ assigned to \gls{vm}s dynamically during the simulation and compare them with the \gls{vm}s' CPU boundedness $\beta$, we counted the number of occurrences of each $(\beta, f)$ combination for every \gls{vm} and time slot. This data is illustrated\
as a bivariate histogram in Fig.~\ref{fig:beta_freq_histogram} with the number of occurrences shown on a logarithmic scale.\
Darker areas indicate a higher number of frequency occurrences for the respective $(\beta, f)$ combination.\
It can be seen that the occurrences of CPU frequencies assigned based on each \gls{vm}'s CPU boundedness match the areas where \gls{vm} prices are high, based on the perceived-performance pricing model from Fig.~\ref{fig:pricing_model}.
The area with high $\beta$ and low $f$, where prices would be the lowest, contains no occurrences.\
The darkest areas of the graph with a high number of occurrences\
represent the balance between energy savings and profit losses,\ 
which is\
in line with the controller requirements\ 
that energy cost savings should be maximised, but not exceeded by revenue losses.

\subsection{Provider Pricing Impact}



\begin{figure}[!b]
\centering
\includegraphics[width=1.0\columnwidth]{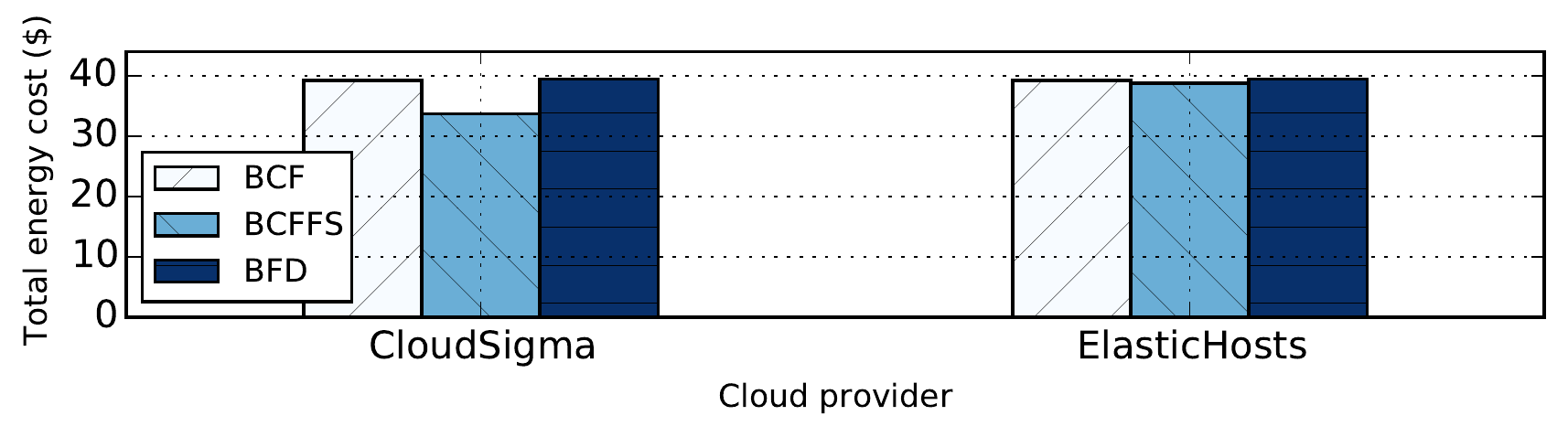}
\caption{Energy costs for the pricing models used by different cloud providers.}
\label{fig:results_provider_pricing}
\end{figure}

In this set of experiments we evaluated and compared the performance of the algorithms for different pricing models in order to investigate the impact of the pricing model on the savings from using the proposed approach. Fig.~\ref{fig:results_provider_pricing} presents the results for the CloudSigma and ElasticHosts cloud providers.\ 
As can be seen, higher energy cost savings are possible for the CloudSigma pricing scheme ($14\%$) than for the pricing offered by ElasticHosts ($2\%$). This is because CPU provisioning offered by CloudSigma is charged at a lower price resulting in service revenue being closer to the energy costs. As a result, energy savings gain comparably more weight in the revenue-energy balancing performed by the cloud controller. 
Since our method applies better to cloud providers like CloudSigma, we used their pricing scheme in all the other simulation scenarios.


\begin{figure}[!b]
\centering
\includegraphics[width=1.0\columnwidth]{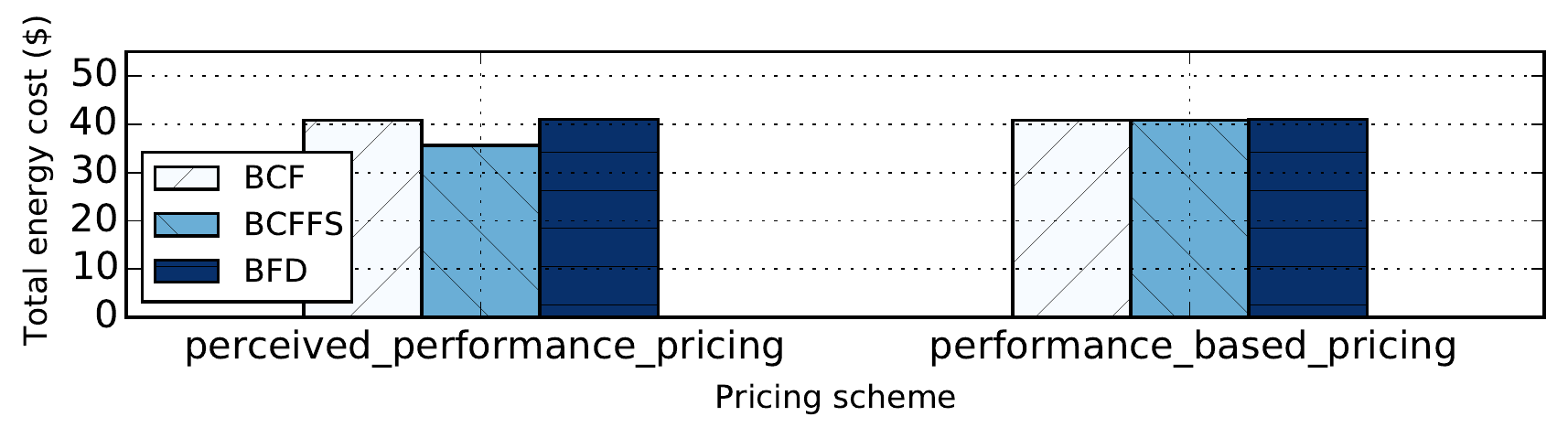}
\caption{Energy cost savings for perceived-performance and performance-based pricing.}
\label{fig:pricing_impact_arm}
\end{figure}

\subsection{Pricing Model Impact}

In this experiment we compared the savings obtained by using different pricing models. These include the perceived-performance pricing model proposed in Section~\ref{section:perc_pricing} and performance-based pricing offered by the current providers. The results\ 
are presented in Fig.~\ref{fig:pricing_impact_arm}.\
It can be seen that using performance-based pricing does not lead to energy savings, as the reduction in prices is high compared with the energy costs. As a result, CPU frequency scaling is not feasible. On the other hand, using perceived-performance pricing, savings are possible as CPU frequency reduction does not lead to substantially reduced service revenues.

\subsection{Electricity Cost Variation}

\begin{figure}[!t]
\centering
\includegraphics[width=1.0\columnwidth]{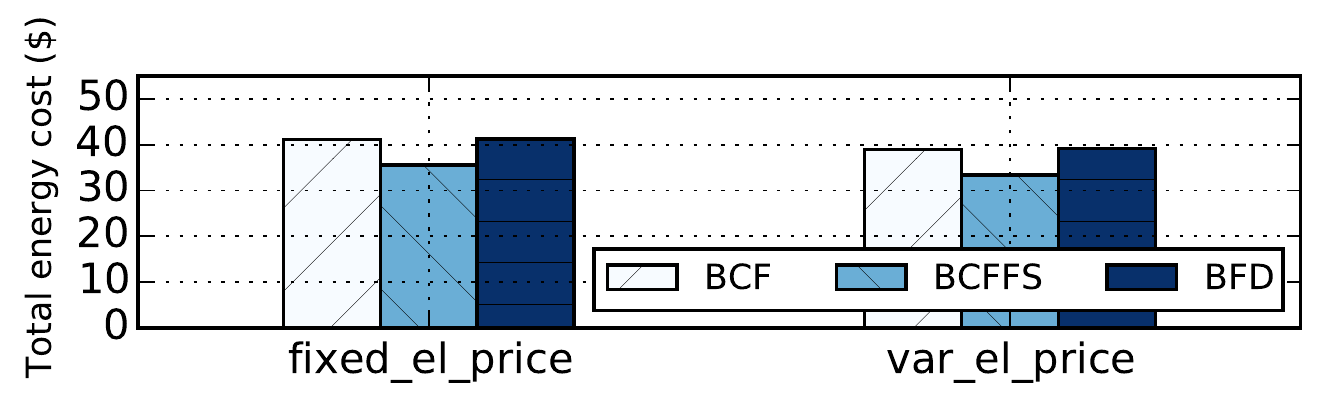}
\caption{Energy cost savings for fixed and variable electricity prices.}
\label{fig:fixed_variable_el_price}
\end{figure}

As not all cloud providers may have access to real-time electricity pricing, in this set of experiments we evaluate the performance of the proposed controller under fixed electricity pricing. In Fig.~\ref{fig:fixed_variable_el_price} scenarios for fixed and variable electricity prices are compared to investigate the impact of electricity pricing on the energy savings obtained using the \schedulerfreq{} controller.\
The \schedulerfreq{} controller achieves better performance under variable electricity pricing reducing the energy costs by exploiting runtime information and adapting the cloud configuration according to the electricity price changes within the day. However, cost savings of $10\%$ (compared to the \schedulerbase{} baseline) that are still significant are achieved for the fixed electricity cost scenario.


\subsection{Variation of Parameter $\beta$}

\begin{figure}[!b]
\centering
\includegraphics[width=1.0\columnwidth]{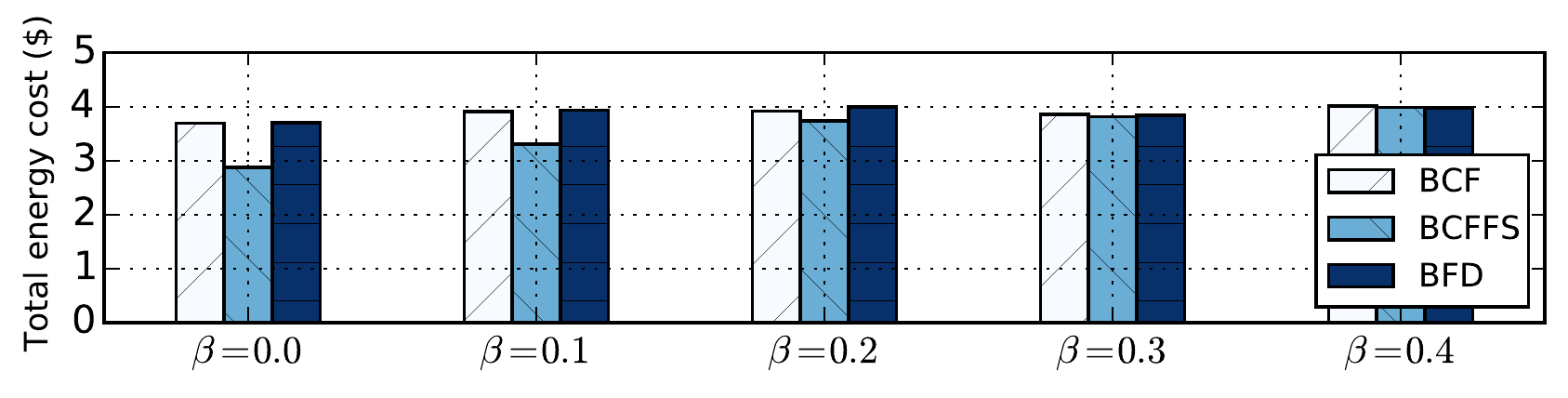}
\caption{Energy cost savings for \gls{vm}s with different fixed CPU-boundedness.}
\label{fig:beta_variation}
\end{figure}

Fig.~\ref{fig:beta_variation} shows the results for \gls{vm}s with fixed CPU-boundedness properties. The aim is to evaluate the impact of different workloads on energy cost savings under the proposed controller and identify workload types where our approach is more beneficial.
To do so, simulations using the same set of \gls{pm}s and \gls{vm} requests were used, while \gls{vm} CPU-boundedness properties were varied between $0.0$ to $0.4$. 
The results are omitted for larger values of $\beta$, where savings are limited due to the impact on application performance.
The energy savings achieved by the \schedulerfreq{} controller decrease gradually while approaching higher values of CPU-boundedness ($\beta$). Between a $\beta$ of 0.0 and 0.2 there is a substantial increase in energy cost as even a small reduction in frequency results in high energy cost savings that exceed the revenue losses. For higher values of $\beta$, the savings are limited due to the impact on application performance. As a result, the \schedulerfreq{} controller achieves the best results for I/O-bound workloads where application performance is not greatly affected by the reduction in frequency.

\section{Conclusion}
\label{sec:conclusion}

As the demand for cloud platforms increases and as the workload becomes more diverse, a one-fit-all pricing policy does not only provide poor flexibility to the user, but is also not energy efficient.
To keep up with the rapid evolution in information infrastructure, a more flexible way of controlling cloud systems must be provided to both satisfy the user and minimize the energy costs.

We have presented a flexible cloud control approach capable of system-level resource management to fit the performance guarantees requested by the user and minimise energy waste by scaling CPU resources on demand.\
Our cloud controller is driven by a 
model which covers realistic aspects of real-world cloud platforms.
Geotemporal inputs such as real-time electricity pricing and temperature-dependent cooling affecting a geographically-distributed cloud provider have been modelled together with a multi-architecture, multi-core power model based on real experiments and used in the Philharmonic cloud simulator to estimate operational costs and the \gls{vm} service revenue. Several scenarios were examined and two baseline methods were used resulting in energy cost savings of up to \ensavingsmax{} for ARM and up to \ensavingsmaxintel{} for Intel architectures.

The lessons learned from our research can be applied to cut costs in data centers. For example, a cloud provider with an \$12M annual electricity bill providing \gls{vm}s on ARM infrastructure at prices similar to CloudSigma for mostly I/O intensive workloads can save around \$1.7M, assuming no frequency scaling was previously used.\
Even if not all ideal factors are satisfied, e.g. the \gls{vm} prices are in the ElasticHosts range, around \$750k savings can be achieved for a larger cloud provider with a \$38M annual energy bill (estimated in~\cite{qureshi_cutting_2009}).

Our results show that energy costs can be significantly reduced without noticeably impacting the service revenue by scaling the CPU frequencies of the \gls{pm}s according to the hosted \gls{vm} characteristics. We have shown that this method applies better to some cloud providers like CloudSigma, where service revenue is closer to energy costs. Savings can be achieved for fixed electricity pricing, but \gls{rtep} pricing allows higher energy savings. For our method, ARM architectures are more suitable than Intel, and more I/O-intensive workloads allow for higher savings than CPU-intensive workloads.

As part of our future work, we would like to investigate approaches where the \gls{vm} migration cloud controller stage also considers the workload CPU-boundedness characteristics\
in order to maximise the energy savings from using perceived-performance pricing.


\ifCLASSOPTIONcompsoc
  \section*{Acknowledgments}
\else
  \section*{Acknowledgment}
\fi

The work described in this paper has been partially funded\
through the projects Haley (Holistic Energy Efficient Hybrid Clouds),\
and NESUS (Network for Sustainable Ultrascale Computing).

\bibliographystyle{IEEEtran-kermit}
\bibliography{references,references-drazen}

\begin{thebibliography}{10}
\providecommand{\url}[1]{#1}
\csname url@samestyle\endcsname
\providecommand{\newblock}{\relax}
\providecommand{\bibinfo}[2]{#2}
\providecommand{\BIBentrySTDinterwordspacing}{\spaceskip=0pt\relax}
\providecommand{\BIBentryALTinterwordstretchfactor}{4}
\providecommand{\BIBentryALTinterwordspacing}{\spaceskip=\fontdimen2\font plus
\BIBentryALTinterwordstretchfactor\fontdimen3\font minus
  \fontdimen4\font\relax}
\providecommand{\BIBforeignlanguage}[2]{{%
\expandafter\ifx\csname l@#1\endcsname\relax
\typeout{** WARNING: IEEEtran.bst: No hyphenation pattern has been}%
\typeout{** loaded for the language `#1'. Using the pattern for}%
\typeout{** the default language instead.}%
\else
\language=\csname l@#1\endcsname
\fi
#2}}
\providecommand{\BIBdecl}{\relax}
\BIBdecl

\bibitem{_gartner_2013}
\BIBentryALTinterwordspacing
``Gartner {Says} {Worldwide} {Public} {Cloud} {Services} {Market} to {Total}
  \$131 {Billion},'' 2013. [Online]. Available:
  \url{http://www.gartner.com/newsroom/id/2352816}
\BIBentrySTDinterwordspacing

\bibitem{elastichosts}
{ElasticHosts}, \url{Available: http://www.elastichosts.co.uk/}.

\bibitem{cloudsigma}
{CloudSigma}, \url{Available: https://www.cloudsigma.com/}.

\bibitem{jonathan_koomey_growth_2011}
\BIBentryALTinterwordspacing
{Jonathan Koomey}, ``Growth in {Data} center electricity use 2005 to 2010,''
  Analytics Press, Oakland, CA, Tech. Rep., Aug. 2011. [Online]. Available:
  \url{http://www.analyticspress.com/datacenters.html}
\BIBentrySTDinterwordspacing

\bibitem{_gartner_2007}
\BIBentryALTinterwordspacing
``Gartner {Estimates} {ICT} {Industry} {Accounts} for 2 {Percent} of {Global}
  {CO}2 {Emissions},'' 2007. [Online]. Available:
  \url{http://www.gartner.com/newsroom/id/503867}
\BIBentrySTDinterwordspacing

\bibitem{weron_modeling_2006}
R.~Weron, \emph{Modeling and {Forecasting} {Electricity} {Loads} and {Prices}:
  {A} {Statistical} {Approach}}, 1st~ed.\hskip 1em plus 0.5em minus 0.4em\relax
  Wiley, Dec. 2006.

\bibitem{xu_temperature_2013}
\BIBentryALTinterwordspacing
H.~Xu, C.~Feng, and B.~Li, ``Temperature aware workload management in
  geo-distributed datacenters,'' in \emph{Proceedings of the {ACM}
  {SIGMETRICS}/international conference on {Measurement} and modeling of
  computer systems}, vol.~41.\hskip 1em plus 0.5em minus 0.4em\relax ACM, 2013,
  pp. 373--374.
\BIBentrySTDinterwordspacing

\bibitem{miyoshi2002critical}
A.~Miyoshi, C.~Lefurgy, E.~Van~Hensbergen, R.~Rajamony, and R.~Rajkumar,
  ``Critical power slope: understanding the runtime effects of frequency
  scaling,'' in \emph{Proceedings of the 16th International Conference on
  Supercomputing (ICS)}.\hskip 1em plus 0.5em minus 0.4em\relax ACM, 2002, pp.
  35--44.

\bibitem{von2009power}
G.~Von~Laszewski, L.~Wang, A.~J. Younge, and X.~He, ``Power-aware scheduling of
  virtual machines in {DVFS}-enabled clusters,'' in \emph{Proceedings of the
  IEEE International Conference on Cluster Computing and Workshops
  (CLUSTER)}.\hskip 1em plus 0.5em minus 0.4em\relax IEEE, 2009, pp. 1--10.

\bibitem{shi2011towards}
W.~Shi and B.~Hong, ``Towards profitable virtual machine placement in the data
  center,'' in \emph{Proceedings of the 4th IEEE International Conference on
  Utility and Cloud Computing}.\hskip 1em plus 0.5em minus 0.4em\relax IEEE,
  2011, pp. 138--145.

\bibitem{holmbacka2015energy}
S.~Holmbacka, S.~Lafond, and J.~Lilius, ``Performance monitor based power
  management for {big.LITTLE} platforms,'' in \emph{Workshop on Energy
  Efficiency with Heterogeneous Computing}, D.~Nikolopoulos and J.-L.
  Nunez-Yanez, Eds.\hskip 1em plus 0.5em minus 0.4em\relax HiPEAC, 2015, p. 1
  – 6.

\bibitem{holmbacka2014energy}
S.~Holmbacka, E.~Nogues, M.~Pelcat, S.~Lafond, and J.~Lilius, ``Energy
  efficiency and performance management of parallel dataflow applications,'' in
  \emph{Conference on Design \& Architectures for Signal \& Image Processing},
  A.~Pinzari and A.~Morawiec, Eds.\hskip 1em plus 0.5em minus 0.4em\relax ECDI
  Electronic Chips \& Systems design initiative, 2014, p. 1 – 8.

\bibitem{rajovic2013supercomputing}
N.~Rajovic, P.~M. Carpenter, I.~Gelado, N.~Puzovic, A.~Ramirez, and M.~Valero,
  ``{Supercomputing with commodity CPUs: are mobile SoCs ready for HPC?}'' in
  \emph{High Performance Computing, Networking, Storage and Analysis (SC), 2013
  International Conference for}.\hskip 1em plus 0.5em minus 0.4em\relax IEEE,
  2013, pp. 1--12.

\bibitem{francesquini2015benchmark}
\BIBentryALTinterwordspacing
E.~Francesquini, M.~Castro, H.~Penna, Pedro, F.~Dupros, C.~D. Freitas,
  Henrique, P.~O.~A. Navaux, and J.-F. Mehaut, ``{On the Energy Eciency and
  Performance of Irregular Application Executions on Multicore, NUMA and
  Manycore Platforms},'' \emph{{Journal of Parallel and Distributed
  Computing}}, vol.~76, pp. pp. 32--48, Feb. 2015.
\BIBentrySTDinterwordspacing

\bibitem{spiliopoulos2014power}
V.~Spiliopoulos, S.~Kaxiras, and G.~Keramidas, ``Green governors: A framework
  for continuously adaptive {DVFS},'' in \emph{Green Computing Conference and
  Workshops (IGCC), 2011 International}, July 2011, pp. 1--8.

\bibitem{etinski2010optimizing}
M.~Etinski, J.~Corbalan, J.~Labarta, and M.~Valero, ``Optimizing job
  performance under a given power constraint in {HPC} centers,'' in
  \emph{Proceedings of the International Green Computing Conference (IGCC)},
  2010, pp. 257--267.

\bibitem{sasaki2013model}
H.~Sasaki, S.~Imamura, and K.~Inoue, ``Coordinated power-performance
  optimization in manycores,'' in \emph{Parallel Architectures and Compilation
  Techniques (PACT), 2013 22nd International Conference on}, 2013, pp. 51--61.

\bibitem{seeker2014energy}
V.~Seeker, P.~Petoumenos, H.~Leather, and B.~Franke, ``Measuring {QoE} of
  interactive workloads and characterising frequency governors on mobile
  devices,'' in \emph{Workload Characterization (IISWC), 2014 IEEE
  International Symposium on}, Oct 2014, pp. 61--70.

\bibitem{lucanin_cloud_2015}
D.~Lucanin, I.~Pietri, I.~Brandic, and R.~Sakellariou, ``A {Cloud} {Controller}
  for {Performance}-{Based} {Pricing},'' in \emph{2015 {IEEE} 8th
  {International} {Conference} on {Cloud} {Computing} ({CLOUD})}, Jun. 2015,
  pp. 155--162.

\bibitem{lucanin2014energy}
D.~Lu{\v{c}}anin, F.~Jrad, I.~Brandic, and A.~Streit, ``Energy-aware cloud
  management through progressive {SLA} specification,'' in \emph{11th
  International Conference on Economics of Grids, Clouds, Systems, and Services
  (GECON)}.\hskip 1em plus 0.5em minus 0.4em\relax Springer, 2014, pp. 83--98.

\bibitem{beloglazov_energy-aware_2012}
\BIBentryALTinterwordspacing
A.~Beloglazov, J.~Abawajy, and R.~Buyya, ``Energy-aware resource allocation
  heuristics for efficient management of data centers for {Cloud} computing,''
  \emph{Future Generation Computer Systems}, vol.~28, no.~5, pp. 755--768, May
  2012.
\BIBentrySTDinterwordspacing

\bibitem{alfeld_toward_2012}
\BIBentryALTinterwordspacing
S.~Alfeld, C.~Barford, and P.~Barford, ``Toward an analytic framework for the
  electrical power grid,'' in \emph{Proceedings of the 3rd {International}
  {Conference} on {Future} {Energy} {Systems}: {Where} {Energy}, {Computing}
  and {Communication} {Meet}}, ser. e-{Energy} '12.\hskip 1em plus 0.5em minus
  0.4em\relax New York, NY, USA: ACM, 2012, pp. 9:1--9:4.
\BIBentrySTDinterwordspacing

\bibitem{planetlab}
{PlanetLab workload traces}, \url{Available:
  https://github.com/beloglazov/planetlab-workload-traces}.

\bibitem{qureshi_cutting_2009}
\BIBentryALTinterwordspacing
A.~Qureshi, R.~Weber, H.~Balakrishnan, J.~Guttag, and B.~Maggs, ``Cutting the
  electric bill for internet-scale systems,'' \emph{SIGCOMM Comput. Commun.
  Rev.}, vol.~39, no.~4, pp. 123--134, Aug. 2009.
\BIBentrySTDinterwordspacing

\bibitem{buchbinder_online_2011}
\BIBentryALTinterwordspacing
N.~Buchbinder, N.~Jain, and I.~Menache, ``Online {Job}-{Migration} for
  {Reducing} the {Electricity} {Bill} in the {Cloud},'' in \emph{{NETWORKING}
  2011}, ser. Lecture {Notes} in {Computer} {Science}, J.~Domingo-Pascual,
  P.~Manzoni, S.~Palazzo, A.~Pont, and C.~Scoglio, Eds.\hskip 1em plus 0.5em
  minus 0.4em\relax Springer Berlin Heidelberg, Jan. 2011, no. 6640, pp.
  172--185.
\BIBentrySTDinterwordspacing

\bibitem{guler_cutting_2013}
H.~Guler, B.~Cambazoglu, and O.~Ozkasap, ``Cutting {Down} the {Energy} {Cost}
  of {Geographically} {Distributed} {Cloud} {Data} {Centers},'' in \emph{Energy
  {Efficiency} in {Large} {Scale} {Distributed} {Systems}}.\hskip 1em plus
  0.5em minus 0.4em\relax Vienna: Springer Berlin Heidelberg, 2013, pp.
  279--286.

\bibitem{liu_renewable_2012}
\BIBentryALTinterwordspacing
Z.~Liu, Y.~Chen, C.~Bash, A.~Wierman, D.~Gmach, Z.~Wang, M.~Marwah, and
  C.~Hyser, ``Renewable and cooling aware workload management for sustainable
  data centers,'' in \emph{Proceedings of the 12th {ACM}
  {SIGMETRICS}/{PERFORMANCE} joint international conference on {Measurement}
  and {Modeling} of {Computer} {Systems}}, ser. {SIGMETRICS} '12.\hskip 1em
  plus 0.5em minus 0.4em\relax New York, NY, USA: ACM, 2012, pp. 175--186.
\BIBentrySTDinterwordspacing

\bibitem{wu2014green}
C.-M. Wu, R.-S. Chang, and H.-Y. Chan, ``A green energy-efficient scheduling
  algorithm using the {DVFS} technique for cloud datacenters,'' \emph{Future
  Generation Computer Systems (FGCS)}, vol.~37, pp. 141--147, 2014.

\bibitem{freeh2007analyzing}
V.~W. Freeh, D.~K. Lowenthal, F.~Pan, N.~Kappiah, R.~Springer, B.~L. Rountree,
  and M.~E. Femal, ``Analyzing the energy-time trade-off in high-performance
  computing applications,'' \emph{IEEE Transactions on Parallel and Distributed
  Systems (TPDS)}, vol.~18, no.~6, pp. 835--848, 2007.

\bibitem{hsuan2013cloud}
C.~hsuan Hsu, C.-C. Lin, and T.-S. Hsu, ``Energy-conscious cloud computing
  adopting {DVFS} and state-switching for workflow applications,'' in
  \emph{Cloud Computing and Big Data (CloudCom-Asia), 2013 International
  Conference on}, Dec 2013, pp. 1--8.

\bibitem{zhuo2014cloud}
Z.~Tang, Z.~Cheng, K.~Li, and K.~Li, ``An efficient energy scheduling algorithm
  for workflow tasks in hybrids and {DVFS}-enabled cloud environment,'' in
  \emph{Parallel Architectures, Algorithms and Programming (PAAP), 2014 Sixth
  International Symposium on}, July 2014, pp. 255--261.

\bibitem{ioannou2011cloud}
N.~Ioannou, M.~Kauschke, M.~Gries, and M.~Cintra, ``Phase-based
  application-driven hierarchical power management on the single-chip cloud
  computer,'' in \emph{Parallel Architectures and Compilation Techniques
  (PACT), 2011 International Conference on}, Oct 2011, pp. 131--142.

\bibitem{alnowiser2014cloud}
A.~Alnowiser, E.~Aldhahri, A.~Alahmadi, and M.~Zhu, ``Enhanced weighted round
  robin {(EWRR) with DVFS} technology in cloud,'' in \emph{Computational
  Science and Computational Intelligence (CSCI), 2014 International Conference
  on}, vol.~1, March 2014, pp. 320--326.

\bibitem{hallis2013power}
F.~Hällis, S.~Holmbacka, W.~Lund, R.~Slotte, S.~Lafond, and J.~Lilius,
  ``Thermal influence on the energy efficiency of workload consolidation in
  many-core architecture,'' in \emph{Proceedings of the 24th Tyrrhenian
  International Workshop on Digital Communications}, R.~Bolla, F.~Davoli,
  P.~Tran-Gia, and T.~T. Anh, Eds.\hskip 1em plus 0.5em minus 0.4em\relax IEEE,
  2013, p. 1–6.

\bibitem{martinez2010model}
\BIBentryALTinterwordspacing
F.~J. Mesa-Martinez, E.~K. Ardestani, and J.~Renau, ``Characterizing processor
  thermal behavior,'' \emph{SIGPLAN Not.}, vol.~45, no.~3, pp. 193--204, Mar.
  2010.
\BIBentrySTDinterwordspacing

\bibitem{rauber2012model}
T.~Rauber and G.~Runger, ``Energy-aware execution of fork-join-based task
  parallelism,'' in \emph{Modeling, Analysis Simulation of Computer and
  Telecommunication Systems (MASCOTS), 2012 IEEE 20th International Symposium
  on}, 2012, pp. 231--240.

\bibitem{tudor2012model}
B.~Tudor and Y.-M. Teo, ``Towards modelling parallelism and energy performance
  of multicore systems,'' in \emph{Parallel and Distributed Processing
  Symposium Workshops PhD Forum (IPDPSW), 2012 IEEE 26th International}, 2012,
  pp. 2526--2529.

\bibitem{cupertino2014model}
\BIBentryALTinterwordspacing
L.~Cupertino, G.~Da~Costa, and J.-M. Pierson,
  ``\BIBforeignlanguage{English}{Towards a generic power estimator},''
  \emph{\BIBforeignlanguage{English}{Computer Science - Research and
  Development}}, pp. 1--9, 2014.
\BIBentrySTDinterwordspacing

\bibitem{shao2013model}
\BIBentryALTinterwordspacing
Y.~S. Shao and D.~Brooks, ``Energy characterization and instruction-level
  energy model of intel's xeon phi processor,'' in \emph{Proceedings of the
  2013 International Symposium on Low Power Electronics and Design}, ser.
  ISLPED '13.\hskip 1em plus 0.5em minus 0.4em\relax Piscataway, NJ, USA: IEEE
  Press, 2013, pp. 389--394.
\BIBentrySTDinterwordspacing

\bibitem{cho2010model}
S.~Cho and R.~Melhem, ``On the interplay of parallelization, program
  performance, and energy consumption,'' \emph{Parallel and Distributed
  Systems, IEEE Transactions on}, vol.~21, no.~3, pp. 342--353, 2010.

\bibitem{shen2012model}
H.~Shen, J.~Lu, and Q.~Qiu, ``Learning based {DVFS} for simultaneous
  temperature, performance and energy management,'' in \emph{Quality Electronic
  Design (ISQED), 2012 13th International Symposium on}, March 2012, pp.
  747--754.

\bibitem{bharathwaj2013model}
B.~Raghunathan, Y.~Turakhia, S.~Garg, and D.~Marculescu, ``Cherry-picking:
  Exploiting process variations in dark-silicon homogeneous chip
  multi-processors,'' in \emph{Design, Automation Test in Europe Conference
  Exhibition (DATE), 2013}, March 2013, pp. 39--44.

\bibitem{kim2003power}
N.~Kim, T.~Austin, D.~Baauw, T.~Mudge, K.~Flautner, J.~Hu, M.~Irwin,
  M.~Kandemir, and V.~Narayanan, ``Leakage current: Moore's law meets static
  power,'' \emph{Computer}, vol.~36, no.~12, pp. 68--75, Dec 2003.

\bibitem{shah2015platform}
\BIBentryALTinterwordspacing
S.~Agam, ``Lenovo building its first prototype {ARM} server,'' February 2015.
\BIBentrySTDinterwordspacing

\bibitem{lawson1987optimization}
C.~L. Lawson and R.~J. Hanson, \emph{Solving Least Squares Problems}.\hskip 1em
  plus 0.5em minus 0.4em\relax Society for Industrial and Applied Mathematics,
  1987.

\bibitem{pietri2015cost}
I.~Pietri and R.~Sakellariou, ``Cost-efficient {CPU} provisioning for
  scientific workflows on clouds,'' in \emph{Proceedings of the 12th
  International Conference on Economics of Grids, Clouds, Systems and
  Services}.\hskip 1em plus 0.5em minus 0.4em\relax Springer LNCS, 2015.

\bibitem{chi2015benchmark}
\BIBentryALTinterwordspacing
C.~C. Chi, M.~Alvarez-Mesa, and B.~Juurlink, ``Low-power high-efficiency video
  decoding using general-purpose processors,'' \emph{ACM Trans. Archit. Code
  Optim.}, vol.~11, no.~4, pp. 56:1--56:25, Jan. 2015.
\BIBentrySTDinterwordspacing

\bibitem{padoin2015benchmark}
E.~Padoin, L.~Lima~Pilla, M.~Castro, F.~Boito, P.~Navaux, and J.-F. Mehaut,
  ``Performance/energy trade-off in scientific computing: the case of {ARM
  big.LITTLE and Intel Sandy Bridge},'' \emph{Computers Digital Techniques,
  IET}, vol.~9, no.~1, pp. 27--35, 2015.

\bibitem{beloglazov_managing_2013}
A.~Beloglazov and R.~Buyya, ``Managing {Overloaded} {Hosts} for {Dynamic}
  {Consolidation} of {Virtual} {Machines} in {Cloud} {Data} {Centers} {Under}
  {Quality} of {Service} {Constraints},'' \emph{IEEE Transactions on Parallel
  and Distributed Systems}, vol.~24, no.~7, pp. 1366--1379, 2013.

\bibitem{drazen_lucanin_philharmonic_2014}
\BIBentryALTinterwordspacing
{Dražen Lučanin}, ``Philharmonic,'' 2014. [Online]. Available:
  \url{https://philharmonic.github.io/}
\BIBentrySTDinterwordspacing

\bibitem{_forecast_2015}
\BIBentryALTinterwordspacing
``Forecast,'' 2015. [Online]. Available: \url{http://forecast.io/}
\BIBentrySTDinterwordspacing

\end{thebibliography}

\begin{IEEEbiography}[{\includegraphics[width=1in,height=1.25in,clip,keepaspectratio]{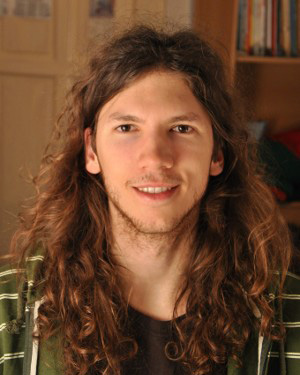}}]{Dražen Lučanin}
is a PhD student at the Vienna University of Technology,\
studying energy efficiency in cloud computing.\
Previously, he worked as an external associate at the Ruđer Bošković Institute\
on machine learning methods for forecasting financial crises.\
He graduated with a master's degree in computer science\
at the Faculty of electrical engineering and computing, University of Zagreb.\
\end{IEEEbiography}

\begin{IEEEbiography}[{\includegraphics[width=1in,height=1.25in,clip,keepaspectratio]{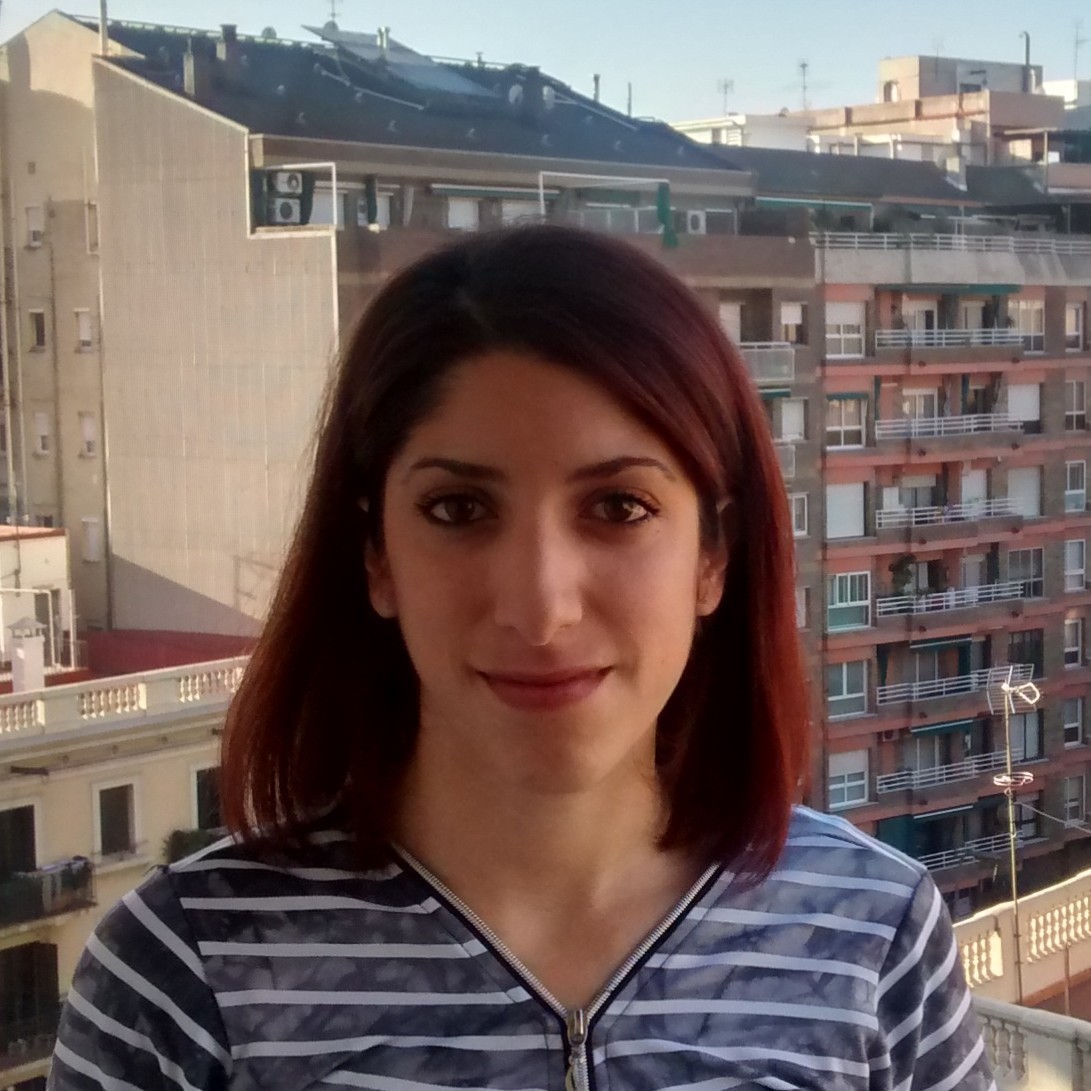}}]{Ilia Pietri}
completed her PhD degree in Computer Science at the University of Manchester, UK, in 2016. She received her BSc degree in Informatics and Telecommunications and MSc degree in Economics and Administration of Telecommunication Networks from the National and Kapodistrian University of Athens, Greece, in 2008 and 2010, respectively. Currently, she is a PostDoc researcher at "Athena" Research Center, Greece. 
Her research interests include resource management and cost efficiency in distributed systems, such as clouds.\
\end{IEEEbiography}

\begin{IEEEbiography}[{\includegraphics[width=1in,height=1.25in,clip,keepaspectratio]{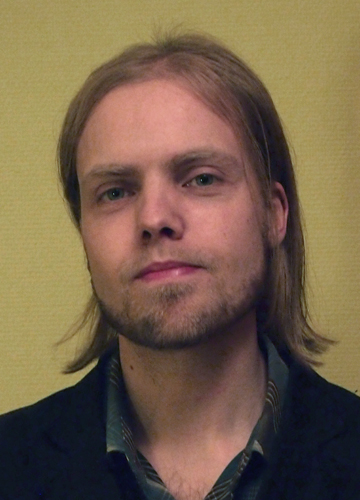}}]{Simon Holmbacka}
received his BSc. and MSc. and PhD degrees in Computer Engineering from Åbo Akademi University of Finland in 2009, 2011 and 2016 respectively. 
He is currently a PostDoc researcher at the Embedded Systems Laboratory at Åbo Akademi. His main research areas are energy efficiency
in many-core systems, ranging from energy aware software to OS level power management, runtime power management and hardware based power management. Other research areas include many-core programming, many-core operating systems, control theory and optimisation.
For more information,\
please visit \url{https://research.it.abo.fi/people/sholmbac}
\end{IEEEbiography}

\begin{IEEEbiography}[{\includegraphics[width=1in,height=1.25in,clip,keepaspectratio]{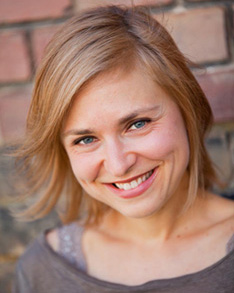}}]{Ivona Brandic}
is Assistant Professor at the Vienna University of Technology. Prior to that, she was Assistant Professor at the Department of Scientific Computing, University of Vienna. She received her PhD degree in 2007 and her venia docendi for practical computer science in 2013, both from Vienna University of Technology. In 2011 she received the Distinguished Young Scientist Award from the Vienna University of Technology for her project on the Holistic Energy Efficient Hybrid Clouds. She published more than 50 scientific journal, magazine and conference publications and she co-authored a text-book on federated and self-manageable Cloud infrastructures. For more information, please visit \url{http://www.ec.tuwien.ac.at/~ivona/}
\end{IEEEbiography}

\begin{IEEEbiography}[{\includegraphics[width=1in,height=1.25in,clip,keepaspectratio]{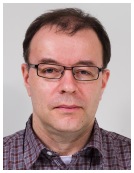}}]{Johan Lilius}
is a full professor of Embedded Systems at Åbo Akademi since 2001. His research interests include Energy-effcient Software, Programming Models for Multicores, System-Level Design, Performance Engineering and Models of Concurrency. He is in charge of the Embedded Systems Curriculum and is teaching several courses in these areas. Prof. Lilius is the author of over 80 publications, has supervised 6 PhD theses and over 50 M.Sc. theses. Prof.
Lilius is also actively participating in the Finnish Strategic Centers for Science, Technology and Innovation in the ICT area, is an associate member of the ARTIST network, and the AAU representative in ARTEMISIA. For more information, please visit \url{https://research.it.abo.fi/people/jolilius}
\end{IEEEbiography}

\begin{IEEEbiography}[{\includegraphics[width=1in,height=1.25in,clip,keepaspectratio]{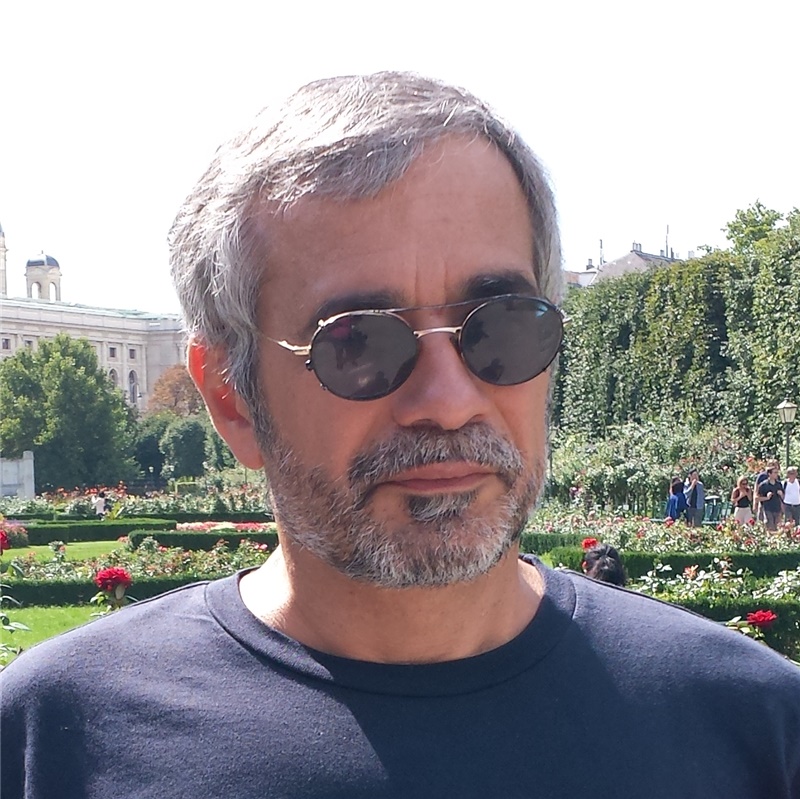}}]{Rizos Sakellariou}
holds a position as a Senior Lecturer (equivalent to Associate Professor in a 3-tier faculty system) with the University of Manchester, UK where he is leading a research laboratory that over the last ten years has hosted more than 25 doctoral students, researchers and visitors. He has carried out research on a number of topics related to parallel and distributed computing and has published over 100 papers in refereed journals and conference proceedings, which have attracted over 3500 Google scholar citations. He has been on the Program Committee of over 110 conferences and he is a member of the Steering Committee of Euro-Par as well as a founding member of the Steering Committee of the newly established Euro-EDUPAR. He values collaboration and a strong work ethic.\
\end{IEEEbiography}

\end{document}